\documentclass[prb,aps,superscriptaddress,twocolumn]{revtex4-1}
\usepackage{amsmath} 
\usepackage{amsthm} 
\usepackage{amssymb}	
\usepackage{graphicx} 
\usepackage[usenames]{color}
\usepackage{array} 
\usepackage{paralist} 
\usepackage{amsfonts}
\usepackage{mathtools}
\usepackage{times}
\usepackage{braket}
\usepackage[colorlinks=true,linkcolor=blue,citecolor=blue,breaklinks]{hyperref}
\normalfont


\renewcommand{\vec}[1]{\ensuremath{\mathbf{#1}}} 
 
\newcommand{\abs}[1]{\left| #1 \right|} 
 
\let\baraccent=\= 
\renewcommand{\=}[1]{\stackrel{#1}{=}} 

\theoremstyle{definition}

\theoremstyle{remark}

\begin{document}
\newcommand\bbone{\ensuremath{\mathbbm{1}}}
\newcommand{\ul}{\underline}
\newcommand{\vl}{v_{_L}}
\newcommand{\vc}{\mathbf}
\newcommand{\be}{\begin{equation}}
\newcommand{\ee}{\end{equation}}
\newcommand{\bk}{{{\bf{k}}}}
\newcommand{\bK}{{{\bf{K}}}}
\newcommand{\cE}{{{\cal E}}}
\newcommand{\bQ}{{{\bf{Q}}}}
\newcommand{\br}{{{\bf{r}}}}
\newcommand{\bg}{{{\bf{g}}}}
\newcommand{\bG}{{{\bf{G}}}}
\newcommand{\hbr}{{\hat{\bf{r}}}}
\newcommand{\bR}{{{\bf{R}}}}
\newcommand{\bq}{{\bf{q}}}
\newcommand{\hx}{{\hat{x}}}
\newcommand{\hy}{{\hat{y}}}
\newcommand{\hd}{{\hat{\delta}}}
\newcommand{\bea}{\begin{eqnarray}}
\newcommand{\eea}{\end{eqnarray}}
\newcommand{\beal}{\begin{align}}
\newcommand{\eeal}{\end{align}}
\newcommand{\ra}{\rangle}
\newcommand{\la}{\langle}
\renewcommand{\tt}{{\tilde{t}}}
\newcommand{\upa}{\uparrow}
\newcommand{\dna}{\downarrow}
\newcommand{\bS}{{\bf S}}
\newcommand{\vS}{\vec{S}}
\newcommand{\dg}{{\dagger}}
\newcommand{\pdg}{{\phantom\dagger}}
\newcommand{\tphi}{{\tilde\phi}}
\newcommand{\cf}{{\cal F}}
\newcommand{\ca}{{\cal A}}
\renewcommand{\ni}{\noindent}
\newcommand{\ct}{{\cal T}}
\newcommand{\zp}[1]{ { \color{red} \footnotesize ZP\; \textsf{\textsl{#1}} } }
\newcommand{\goto}{{\rightarrow}}

\newcommand{\cn}[2]{#1_{#2}^\dagger}
\newcommand{\an}[2]{#1_{#2}^{\phantom\dagger}}
\newcommand{\inv}{{\operatorname{-1}}}

\title{Emergence of Chiral Spin Liquids via Quantum Melting of Non-Coplanar Magnetic Orders}

\author{Ciar\'{a}n Hickey}
\affiliation{Department of Physics, University of Toronto, Toronto, Ontario M5S 1A7, Canada}

\author{Lukasz Cincio}
\affiliation{Theoretical Division, Los Alamos National Laboratory, Los Alamos, NM 87545, USA}
\affiliation{Perimeter Institute for Theoretical Physics, Waterloo, Ontario N2L 2Y5, Canada}

\author{Zlatko Papi\'{c}}
\affiliation{School of Physics and Astronomy, University of Leeds, Leeds, LS2 9JT, United Kingdom}

\author{Arun Paramekanti}
\affiliation{Department of Physics, University of Toronto, Toronto, Ontario M5S 1A7, Canada}
\affiliation{Canadian Institute for Advanced Research, Toronto, Ontario M5G 1Z8, Canada}

\begin{abstract}
Quantum spin liquids (QSLs) are long-range entangled states of quantum magnets which lie beyond the Landau paradigm of classifying phases of matter via broken symmetries. A physical route to arriving at QSLs is via frustration-induced quantum melting of ordered states such as valence bond crystals or magnetic orders. Here, we show, using extensive exact diagonalization (ED) and density-matrix renormalization group (DMRG) studies of concrete $SU(2)$ invariant spin models on honeycomb, triangular and square lattices, that chiral 
spin liquids (CSLs) emerge as descendants of triple-$Q$ spin crystals with tetrahedral magnetic order and a large scalar spin chirality. Such ordered-to-CSL melting transitions 
may yield lattice realizations of effective Chern-Simons-Higgs field theories. Our work provides a distinct unifying perspective on the emergence of CSLs, and suggests that 
materials with magnetic skyrmion crystal order might provide a good starting point to search for CSLs.
\end{abstract}
\maketitle

Quantum spin liquids (QSLs) are phases of matter which defy a classical Landau description in terms of broken symmetries and local order parameters. \cite{QSLRMP2017,Savary2017} Unlike ordered phases, which can be described using simple mean field wavefunctions with short-range entanglement, 
QSLs feature long-range entanglement and unusual excitations with fractional quantum numbers. This leads to robust many-body properties of QSLs 
which are of potential use in topological quantum memories and quantum computation. \cite{NayakRMP2008}

In this paper, we focus on two-dimensional (2D) chiral spin liquids (CSLs), close cousins of the celebrated fractional quantum Hall states. CSLs 
exhibit topological ground state degeneracies, possess gapped anyonic excitations, and 
were originally proposed by Kalmeyer and Laughlin in 
1987 \cite{KalmeyerLaughlin} as candidate ground states of the spin-$1/2$ triangular lattice Heisenberg antiferromagnet (although this model
Hamiltonian is now known to have long-ranged magnetic order). Specifically, viewing the spins as hard-core bosons, the Kalmeyer-Laughlin state
is equivalent to a $\nu\!=\! 1/2$ bosonic Laughlin liquid with gapped semion excitations.

The surge of recent interest in such CSLs
started with the introduction of parent Hamiltonians or exactly solvable models, \cite{SchroeterCSL2007,ThomaleCSL2009,ThomaleNACSL2014,Meng2015} as well as numerical studies of a
variety of simple frustrated spin models which yielded CSL ground states on the kagome, \cite{YHe2014,Bauer2014, Gong2014, FradkinCSL2014,Gong2015, Wietek2015,DNShengKagome2015,ShengVMC2015,BieriVWFKag2015,FradkinCSL2015} square, \cite{Nielsen2013,Poilblanc2015,XJLiu2016} honeycomb, \cite{Hickey2016}
and triangular lattices. \cite{ShengVMC2016,Wietek2016}
CSLs have also been described using variational Gutzwiller projected fermion or boson wavefunctions, whose low energy
properties are captured in terms of spin-$1/2$ partons (spinons) coupled to emergent dynamical gauge fields.
From this perspective, we can obtain the CSL ground state by starting with spin-$1/2$ fermions filling up individual Chern bands with $C=1$, 
leading to an integer quantum Hall state with $\sigma^{\rm total}_{xy}\!=\! 2e^2/h$, and Gutzwiller projecting this state (which
enforces the constraint of one fermion per site) to yield a legitimate spin wavefunction. \cite{WenCSL1989,WenCSL1991,Zhang2011, Zhang2012} 
A complementary picture is to view them as Gutzwiller projected integer quantum Hall states of strongly interacting
bosons with $\sigma_{xy}\!=\! 2e^2/h$. \cite{Barkeshli2013,YCHe2015}
Gutzwiller projection promotes the global $U(1)$ symmetry of the fermions or bosons to a local gauge 
invariance, leading to an emergent low energy $U(1)_2$ Chern Simons theory.

A physically different way to think about QSLs is to start from broken symmetry phases of $SU(2)$ magnets, and introducing strong quantum fluctuations 
to melt the long-range order. For instance, certain frustrated quantum magnets support ordered crystals of valence bond singlets between nearby spins. 
However, quantum fluctuations of such singlet dimers can melt the crystalline order, resulting in a quantum superposition of dimer configurations, which 
provides the resonating valence bond description of gapped $Z_2$ QSLs. \cite{Moessner01} We can also arrive at such a QSL by quantum disordering a coplanar spin spiral, 
without simultaneously proliferating $Z_2$ vortices which are topological point defects in the magnetically ordered phase. \cite{chubukov94}

Here, we focus on the question of how to realize CSLs from quantum disordering magnetically ordered states. We consider previously
discovered $SU(2)$ invariant CSLs on honeycomb, triangular and square lattices, and use extensive 
numerical exact diagonalization (ED) and density matrix renormalization group (DMRG) calculations to show that they descend from
parent non-coplanar
magnetic orders which are skyrmion crystals with zero net magnetization and a nonzero scalar spin chirality. Our work thus presents a distinct
unifying perspective on such chiral spin liquids, and suggests that Mott insulators with magnetic skyrmion crystal order might be viable
candidates for realizing chiral spin liquids - by tuning exchange couplings via physical pressure or chemical composition in order to
melt the  magnetic order.

Our study relies crucially on the classification of so-called ``regular magnetic orders'' (RMOs): magnetically ordered states which preserve all lattice 
symmetries modulo
global spin rotations. \cite{Misguich11} (We note that the original classification of classical RMOs 
considered orders which preserved all lattice symmetries up to global spin rotations and spin-inversion $\vec S \to - \vec S$. However, spin inversion 
is not a symmetry for quantum spins or even a classical symmetry in the presence of chiral interactions, so we drop spin inversion in our
definition of RMOs.)
If we start from such a RMO, and introduce strong quantum fluctuations and frustration, we might expect to restore
a fully symmetric liquid state of spins. This suggests that RMOs with non-coplanar order, net zero magnetization, 
and a large scalar spin chirality are likely to be natural candidates for parent states of singlet
CSLs.

As an example, we have recently studied the phase diagram of an extended Heisenberg model on the honeycomb lattice with additional chiral spin interactions, \cite{Hickey2016} 
and found that the CSL emerges in proximity to a tetrahedral state, which is a non-coplanar RMO. Similar results were subsequently found in a triangular lattice spin
model. \cite{Wietek2016} Here, we
present careful ED and DMRG calculations of the fidelity and energy in these two models, which show that the tetrahedral state and the CSL appear to be separated by a continuous
rather than a first-order transition. This signifies that the CSL may be viewed as descending from the tetrahedral state on both lattices. Interestingly, such tetrahedral
orders and the possibility of topological phases arising from them was pointed out in previous work on itinerant fermion 
models.\cite{Martin2008,Kato2010, Rahmani2013,YRan2014}

We next turn to the square lattice,
for which there are {\it no} non-coplanar RMOs (as we define it).
However, if we allow for $C_4$ symmetry breaking, it turns out there is a non-coplanar RMO
with net zero magnetization, which is a distorted `tetrahedral umbrella' state.
We therefore focus here on Hamiltonians on the square lattice which explicitly break the $C_4$ symmetry of the Hamiltonian, by including a staggered chiral interaction.
This does not impact the topological order of the CSL, since it should survive even in the presence of such symmetry breaking.
However, the simplification is that at the transition from the magnetically ordered state to the CSL, we only need to restore spin rotation symmetry while inducing topological
order, similar to the triangular and honeycomb lattice examples. Indeed, we find that in this case, the combination of staggered chiral
interaction and a further neighbor Heisenberg coupling again drives what appears to be a continuous quantum phase transition between a tetrahedral umbrella
state and a square lattice CSL.

We conjecture that the phases and phase transitions we have uncovered in our numerical studies provide concrete microscopic realizations of 
Chern-Simons-Higgs field theories. In this scenario, the
CSL is the phase with gapped matter fields (bosonic spinons) minimally coupled to a $U(1)_2$ Chern Simons gauge theory in its deconfined phase, 
while the ordered phase is a Higgs condensate of spinons
which leads to magnetic order and a simultaneous loss of topological order.

\section{Model Hamiltonians}

The models we study in this paper are $SU(2)$ invariant spin Hamiltonians with extended Heisenberg interactions, supplemented by a chiral $3$-spin
interaction:
\be
H_{\rm spin} = \frac{1}{2} \sum_{i,j} J_{i,j} \vec S_i \cdot \vec S_j + J_\chi   \sum_{i,j,k \in \triangle} \vec S_i\cdot ( \vec S_j  \times \vec S_k ).
\ee

\begin{figure}[b]
\includegraphics[scale=0.25]{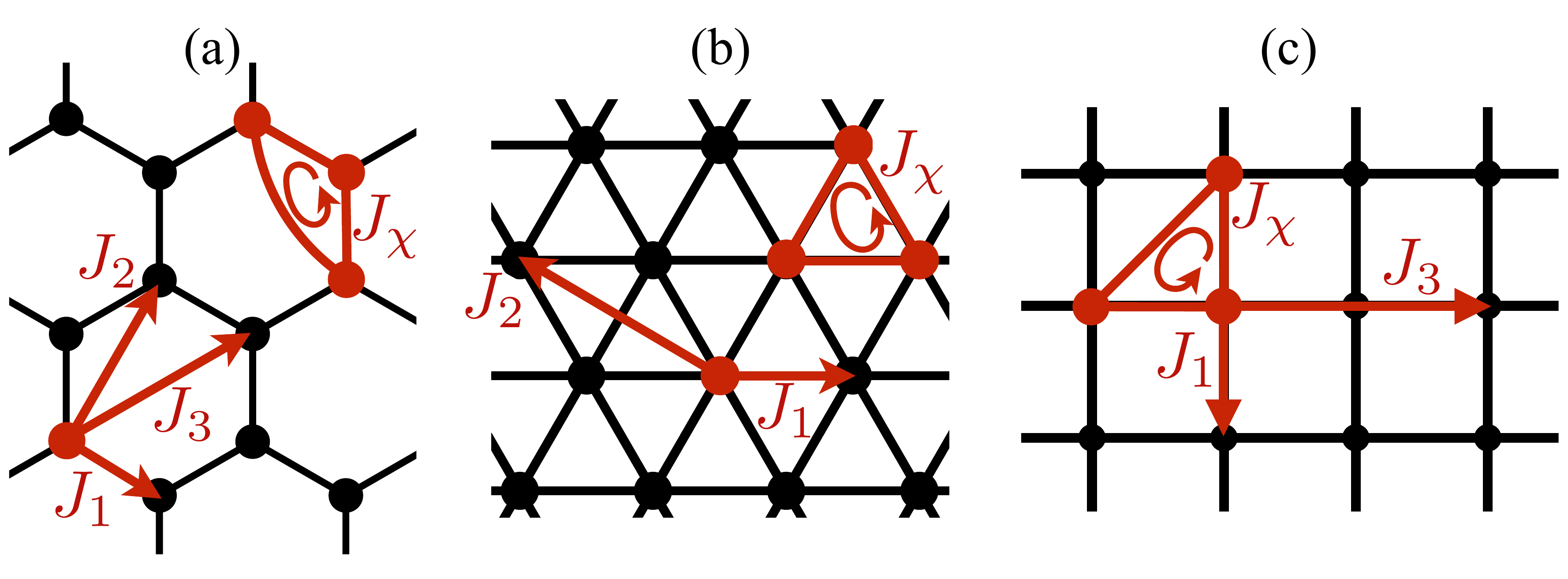}
\caption{Heisenberg exchange $J_{\alpha}$ and scalar spin chirality $J_\chi$ interactions used in the text for the (a) honeycomb, (b) triangular and (c) square lattice. }
\label{fig:Lattice}
\end{figure}

Here, as indicated in Fig.~\ref{fig:Lattice}, the extended Heisenberg interactions include the first few neighbors, while $\sum_\triangle$ in the chiral interaction
denotes a sum over the smallest triangular plaquettes, with $\{i,j,k\}$ taken
anti-clockwise around the triangle.

We emphasize that the chiral terms, which explicitly break time-reversal symmetry, are not necessarily crucial for realizing CSLs; indeed, it has been shown on the
kagome lattice that a CSL with spontaneous breaking of time-reversal symmetry can be realized in an extended Heisenberg model. \cite{Gong2015} Nevertheless, 
we keep these terms since models including such chiral terms have been shown to realize 
CSLs on most 2D lattices - honeycomb, triangular, square, and kagome - and our main aim here is to relate these CSLs to underlying
parent magnetic orders.

If we start from a Hubbard model (with hoppings $t_{i,j}$ and local repulsion $U$), and attempt to derive $H_{\rm spin}$ as an effective spin Hamiltonian 
in the Mott limit, we find $J_{i,j}=4 t^2_{ij}/U$, while the chiral interaction is $J_\chi = 24 (t^3/U^2) \sin\Phi_\triangle
$ if the Hubbard model has
nonzero orbital fluxes $\Phi_\triangle$ penetrating the triangular plaquettes. 
Such chiral terms have been suggested to be relevant for understanding orbital magnetic field effects in certain organic
spin liquids, \cite{MotrunichOMF2006} and for interacting ultracold atomic systems with `artificial' gauge fields (where models such as the honeycomb lattice
Haldane model of a quantum anomalous Hall insulator have been experimentally realized \cite{Haldane1988,Jotzu2014}). 
More recently, there has been a very interesting
suggestion to induce such chiral terms via circularly polarized light rather than by an orbital magnetic field. \cite{Devereaux2016,Aoki2017}
The Hubbard model derivation suggests that the chiral terms will greatly influence magnetism in the `weak' Mott insulator 
regime $U \gtrsim t$. However, as $U/t \to \infty$, i.e. deep in the Mott insulator regime, the chiral terms will become less important than two-spin interactions. 
In our study we will treat this Hamiltonian $H_{\rm spin}$ simply
as a spin model in its own right.

\section{Numerical Signatures}

Within an ED calculation, a CSL and a magnetically ordered state can be distinguished by their low-lying energy spectrum and the static spin structure factor of their ground state. In a magnetically ordered state, with net zero magnetization, the first excited state should be a spin triplet with momenta associated with the magnetic ordering wavevectors. It 
forms part of the ``Anderson tower of states" that 
lead to a symmetry broken ground state in the $N\rightarrow \infty$ limit. \cite{LhuillierQDJS1992,LhuillierSpectra1994} The static spin structure factor of the ground state, $\mathcal{S}(\vec{q})=\frac{1}{N}\sum_{i,j} \braket{\vec{S}_i \cdot \vec{S}_j} e^{i\vec{q}\cdot(\vec{r}_i - \vec{r}_j)} $, 
should display clear peaks at the expected ordering wavevectors. 
On the other hand, in the CSL phase, both the ground state and the first excited state should be spin singlet states with momentum $\bk=(0,0)$ on all of the lattices studied here ($L_x \times L_y$ lattices with both $L_x$ and $L_y$ even, and periodic boundary conditions). The gap between these two states should vanish in the $N \to \infty$ limit, so that they form the degenerate ground state manifold (GSM) of the topologically ordered CSL on the torus. The static spin structure factor of the ground state should be featureless, with no sharp peaks, indicative of a magnetically disordered state.

Our DMRG calculations are done in the infinite cylinder limit with varying circumferences $L_y$. Within such DMRG calculations, one indicator that we are in a CSL 
is that we are able to find multiple distinct ground states associated with the topological order. By contrast, in regimes where we expect magnetic order, we always find only 
a single ground state. A more positive indicator is to calculate the total quantum dimension which is encoded in the overlap of a single ground state on a torus $|\Psi_i\ra$ with its
rotated version, i.e., $R_{ii} \!=\! \bra{\Psi_i} {\mathcal R}_{\pi/3} \ket{\Psi_i}$. Here, ${\mathcal R}_{\pi/3}$ denotes $\pi/3$ rotation of a state on a torus (the appropriate rotation for a lattice with 
$C_6$ symmetry). In an Abelian topologically ordered phase 
with total quantum dimension $D$, we expect $|R_{ii}| = 1/D$,
with $D=\sqrt{2}$ for a CSL. In a topologically trivial state, however, we expect $|R_{ii}| =1$. As a more complete characterization, we can also study the $S$ and $T$
matrices of the anyons in the topologically ordered state by constructing the matrix $R_{ij} = \bra{\Psi_i} {\mathcal R}_{\pi/3} \ket{\Psi_j}$ of overlaps between all pairs of states in the
GSM. \cite{Cincio2013} However, this is not something that we can track across the transition. Yet another distinction between the CSL and magnetically ordered states lies in the edge 
entanglement spectrum obtained by cutting the cylinder into two halves, with the CSL showing a signature of a free chiral boson
described by an $SU(2)_1$ Wess-Zumino-Witten (WZW) conformal field theory. \cite{Cincio2013}

To probe the nature of the transition between the CSL and the magnetically ordered states,
we calculate the ground state energy per spin and the ground state fidelity, usually defined as $F(g)\!=\! |\la\Psi(g)|{\Psi(g+\delta g)}\ra|$, where the ground state $|\Psi (g)\ra$ is parameterized by
the tuning parameter $g$. The fidelity has
been shown to be an indicator of quantum phase transitions, both symmetry-breaking transitions as well as certain topological phase transitions. \cite{Fidelity2006,FidelityReview2010} 
A first order transition is signalled by a sharp discontinuity in $F$, which jumps to zero at the transition where there is a ground state level crossing. By contrast,
a continuous transition is signalled by a weak dip in $F$ at the transition, which results in a (more clearly visible) peak in the fidelity susceptibility 
$\chi^F(g) = \partial^2 F(g)/\partial g^2$.
In the case of topologically ordered states the ground state is not unique so we instead define a GSM fidelity. If 
the states in the GSM do not mix with one another and there are no exact degeneracies (both of which are conditions satisfied in the models and clusters studied here
using ED), 
then we can define the GSM fidelity for an $n$-fold degenerate GSM as
\begin{align}
F_n(g) =  \frac{1}{n} \sum_{i=1}^n \abs{\braket{\Psi_i(g)|\Psi_i(g+\delta g)}} \end{align}
where $n=2$ for the CSL. Associated with this, we define the fidelity susceptibility $\chi^F_n = \partial^2 F_n/\partial g^2$, which we
compute as the numerical second derivative of the fidelity
\begin{equation}
\chi^F_n = \frac{F_n(g+\delta g) - 2 F_n(g) + F_n(g-\delta g)}{(\delta g)^2}.
\end{equation}
In ED, we study $F_2$ and $\chi^F_2$ since we can track the two states that make up the GSM in the 
CSL, GS$1$ and GS$2$, throughout the phase diagram, i.e., we can adiabatically follow the two CSL ground states even into 
the topologically trivial phase (where the upper of the two states in the GSM levitates into a genuine excited state).
In the DMRG calculations, however, we do not have access to this second state in the topologically trivial phase. We thus compute 
the fidelity $F$ of just a single GS; in the CSL, this corresponds to tracking that ground state which is adiabatically connected to the
unique ground state we find in the topologically trivial phase.

\section{Results}

Below, we discuss the results we obtain from both ED and DMRG studies for the honeycomb, triangular and square lattices. We defer a 
discussion of the kagome lattice CSL to a future publication. The data presented is from the largest system sizes studied. For ED this is $N=32$ sites for the 
honeycomb lattice and $N=36$ sites for the triangular and square lattices. In the DMRG simulations we studied infinite cylinders of width up to 6 lattice unit cells for honeycomb and triangular lattices. We keep at most $\chi=2048$ states in the infinite DMRG algorithm, finding convergence in all quantities of interest. $\chi$ is referred to as bond dimension throughout the remaining part of text.
Results on smaller clusters are consistent with 
the conclusions presented here.  

\subsection{Honeycomb Lattice}

\begin{figure}[tb]
\includegraphics[scale=0.25]{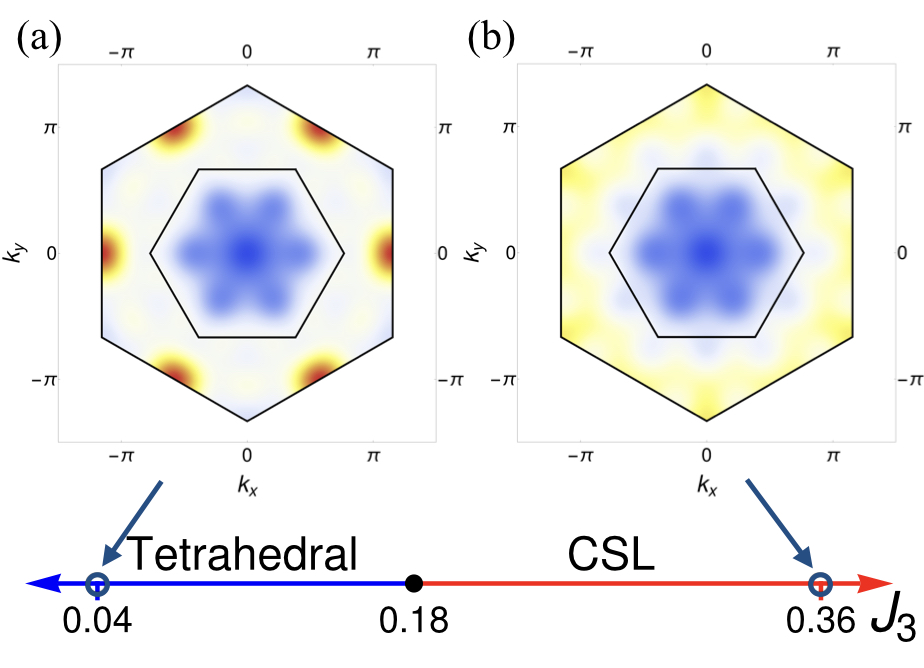}
\caption{Phase diagram of the honeycomb lattice model for $N=36$ sites 
as a function of $J_3$ for fixed $J_1=1.0$, $J_2=0.36$, $J_\chi=0.31$. The static spin structure factor, $S(\vec q)$, of 
the ED ground state for $N=32$ sites is shown for the (a) tetrahedral and (b) CSL phases. $S(\vec q)$ exhibits sharp peaks at
the $M$-points of the (second) BZ in the tetrahedral state, but not in the CSL state.}
\label{fig:HCPD}
\end{figure}

For the honeycomb lattice, we keep Heisenberg terms, $J_1,J_2,J_3$ corresponding to first, second, and third nearest neighbors, in addition to the chiral term $J_{\chi}$,
as shown in Fig.~\ref{fig:Lattice}(a). Our recent study of this model found a tetrahedral phase in this model for $J_3=0$, and a CSL phase for sufficiently large $J_3>0$.
Here, we fix $J_1=1.0$, $J_2=0.36$ and $J_\chi=0.31$, and use ED and DMRG to investigate the nature of the transition between the tetrahedral state and the
CSL state upon increasing $J_3$.

\begin{figure}[tb]
\includegraphics[scale=0.25]{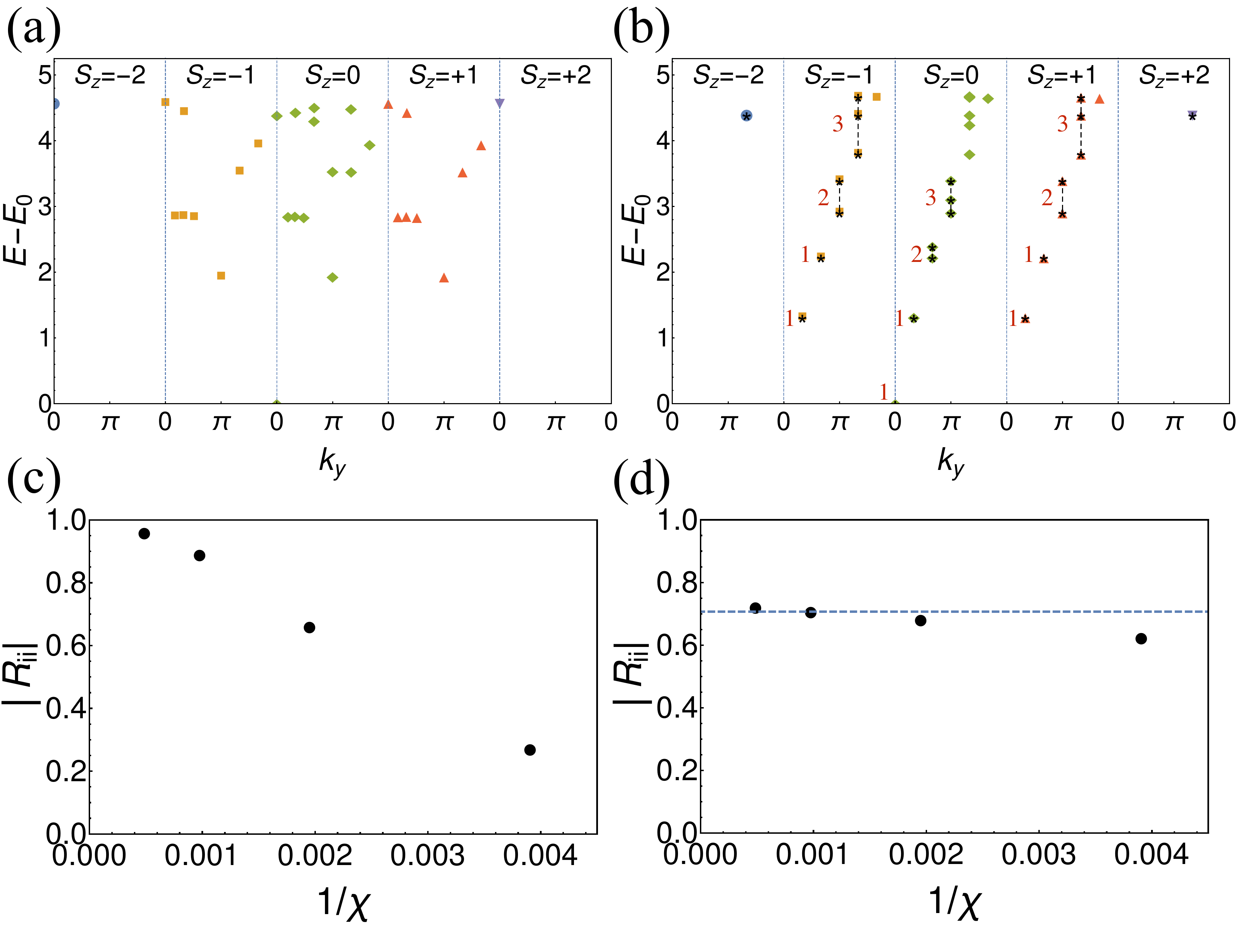}
\caption{The edge entanglement spectrum
obtained using DMRG for the honeycomb model is shown in the (a) tetrahedral ground state,
and (b) one of the two CSL ground states showing the free chiral boson spectrum. The 
overlap $|R_{ii}| \!=\! |\bra{\Psi_i} {\mathcal R}_{\pi/3} \ket{\Psi_i}|$ for various bond dimensions $\chi$ is 
shown in (c) the tetrahedral phase, where it extrapolates to $|R_{ii}| =1$, and (d) the CSL phase, 
where it extrapolates to $|R_{ii}| =1/\sqrt{2}$ (the dashed blue line).}
\label{fig:HCES}
\end{figure}

\begin{figure}[tb]
\includegraphics[scale=0.2]{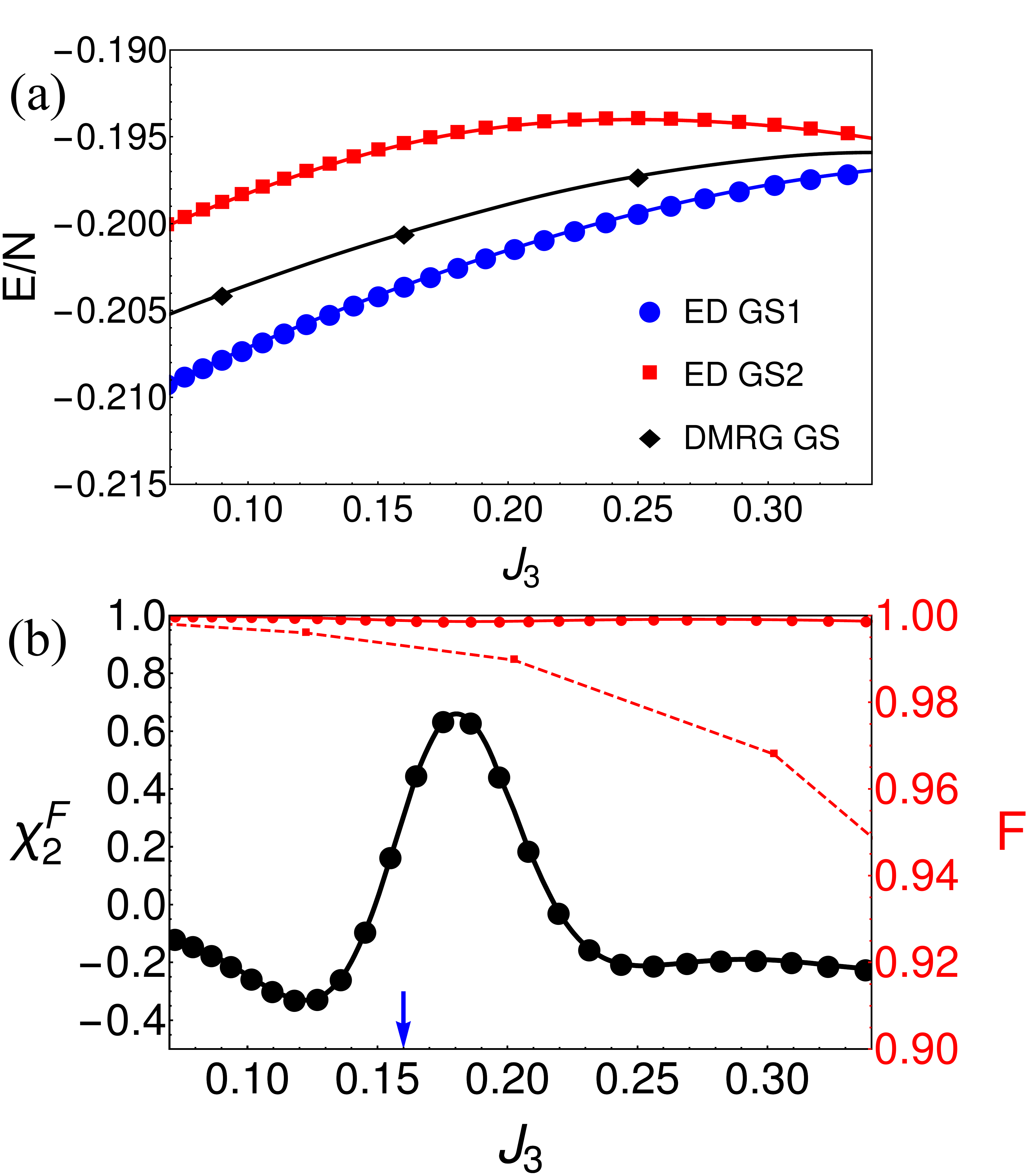}
\caption{(a) Energy per site for the honeycomb model of a single ground state from DMRG and of the two states from ED that make up the ground state manifold of the CSL.
(b) Fidelity of a single ground state from DMRG (dashed red line) and of the ground state manifold from ED (solid red line). The decrease in the DMRG 
fidelity with increasing $J_3$ is due to the nearby CSL-N\'eel transition at even larger $J_3$. The fidelity susceptibility from ED (solid black line) 
shows a clear peak at $J_3=0.18$, signifying the transition from the tetrahedral to CSL. The arrow indicates the position at which the first excited state 
has a level crossing, switching from a triplet at the $M$ point in the tetrahedral phase to a singlet at the $\Gamma$ point in the CSL. }
\label{fig:HCFid}
\end{figure}

In ED, for small $J_3 \geq 0$, we find that the ground state is a spin singlet, 
with large uniform ground state spin chirality $\la \vec S_i \cdot \vec S_j \times \vec S_k \ra \sim 0.16$ on the small triangles of the 
honeycomb lattice (shown in Fig.~\ref{fig:Lattice}(a)) consistent with strong non-coplanarity of the spins.
The first excited state is a spin triplet with momenta at the $M$ points, the ordering wavevector of the tetrahedral state,
and the static spin structure factor,
shown in Fig. \ref{fig:HCPD}(a) for $J_3=0.04$, also exhibits clear peaks at the $M$ points. 
The full spectrum, shown in our previous study (see Supplemental Material of 
Ref.~\onlinecite{Hickey2016}), exhibits an `Anderson tower' consistent with a non-coplanar ground state having fully broken spin rotational symmetry.
All these are clear signatures of the triple-$Q$ tetrahedral state.

By contrast, at 
$J_3=0.36$, we find that the system is in the CSL phase. At this point, ED shows that both the ground state and the first excited state are spin singlet states with 
momentum $\bk\!=\! (0,0)$, and they are well separated by a gap from all other excitations. \cite{Hickey2016} 
This is consistent with a two-fold topological ground state degeneracy on the torus
in the thermodynamic limit, and we have confirmed this by computing the many-body Chern number using flux-threading. \cite{Hickey2016} 
In addition, the static spin structure factor, shown in Fig. \ref{fig:HCPD}(b), exhibits no sharp peaks, indicating short-ranged 
spin correlations.

As a further indication that there is a topological phase transition, DMRG always finds two ground states
at $J_3=0.36$. For either
ground state $\ket{\Psi_i}$, the overlap 
$\abs{\bra{\Psi} {\mathcal R}_{\pi/3} \ket{\Psi}} \approx 1/\sqrt{2}$ for various bond dimensions as shown in
Fig.~\ref{fig:HCES}(d). This indicates that it is consistent with a topologically 
ordered CSL having total quantum dimension $D=\sqrt{2}$. 
Fig.~\ref{fig:HCES}(b) shows the DMRG edge entanglement spectra in the CSL phase
(here, we have picked one of the two ground states), which clearly resembles that of a free chiral boson.
In our previous study of this model, we have provided further evidence for the CSL phase, including the full $S$ and $T$ matrices.
We contrast these observations with the results at $J_3=0.04$. Here, we only find a single ground state, with the overlap 
$\abs{\bra{\Psi} {\mathcal R}_{\pi/3} \ket{\Psi}}$ extrapolating to $1$ with 
increasing bond dimension $\chi$ as shown in Fig.~\ref{fig:HCES}(c), indicative of a topologically trivial phase.
Upon decreasing $J_3$, and entering the tetrahedral state, we find many
additional low-lying states in the entanglement spectrum as seen from Fig.~\ref{fig:HCES}(a), leading to a complete breakdown of the free chiral boson description.

Taken together, these results provide clear evidence that, somewhere between $J_3=0.04$ and $J_3=0.36$, there must be a phase transition 
between a magnetically ordered tetrahedral phase and the CSL. In order to study this transition, we have computed the energy per site as well as the fidelity and
fidelity susceptibilities with varying $J_3$. As shown in 
Fig. \ref{fig:HCFid}(a), the energy shows no sharp kinks, suggesting that this transition is not obviously first order. Similarly the fidelity of the ground states from ED and DMRG,
shown in Fig.~\ref{fig:HCFid}(b) are 
smooth, again suggesting a continuous transition. $\chi^F_2$, computed using ED, has a clear peak at $J_3=0.18$, which we take to be the transition point for the
$N=32$ site system. 


In summary, our ED and DMRG results indicate that the tetrahedral state may be viewed as a parent magnetically ordered state from which the CSL descends
via a seemingly continuous transition upon increasing frustration through $J_3>0$.

\subsection{Triangular Lattice}

For the triangular lattice, we consider a model with nearest and next-nearest neighbor Heisenberg terms $J_1,J_2$, supplemented by a nonzero $J_\chi$. 
For fixed $J_1=1.0$, the phase diagram of the model has already been determined using ED (with up to $36$ spins) over a range of couplings
$0 \leq J_2 \leq 0.3$ and $0 \leq J_\chi \leq 0.6$.
Among other phases, a CSL phase was found bordering a magnetically ordered tetrahedral phase. Here, we fix $J_\chi=0.4$, and vary $J_2$ in
order to investigate the nature of the transition from the CSL to the tetrahedral.

\begin{figure}[tb]
\includegraphics[scale=0.25]{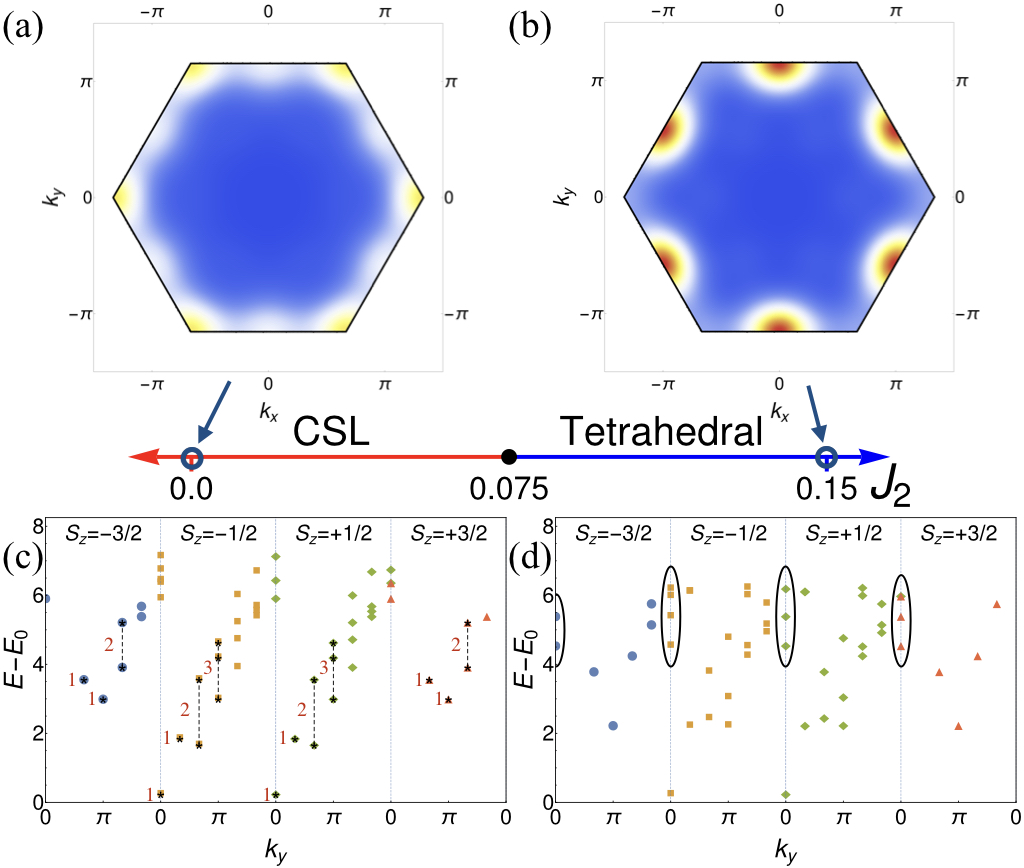}
\caption{Phase diagram of the triangular lattice model for $N=36$ sites as a function of $J_2$ for fixed $J_1=1.0$, $J_\chi=0.4$. The static spin structure factor of 
the ED ground state for $N=36$ spins is shown for the (a) CSL and (b) tetrahedral phases. The DMRG edge entanglement spectra in (c) the CSL phase, showing 
the free chiral boson spectrum, and (d) the tetrahedral phase, where the additional low-lying states are highlighted. }
\label{fig:TriPD}
\end{figure}


\begin{figure}[tb]
\includegraphics[scale=0.2]{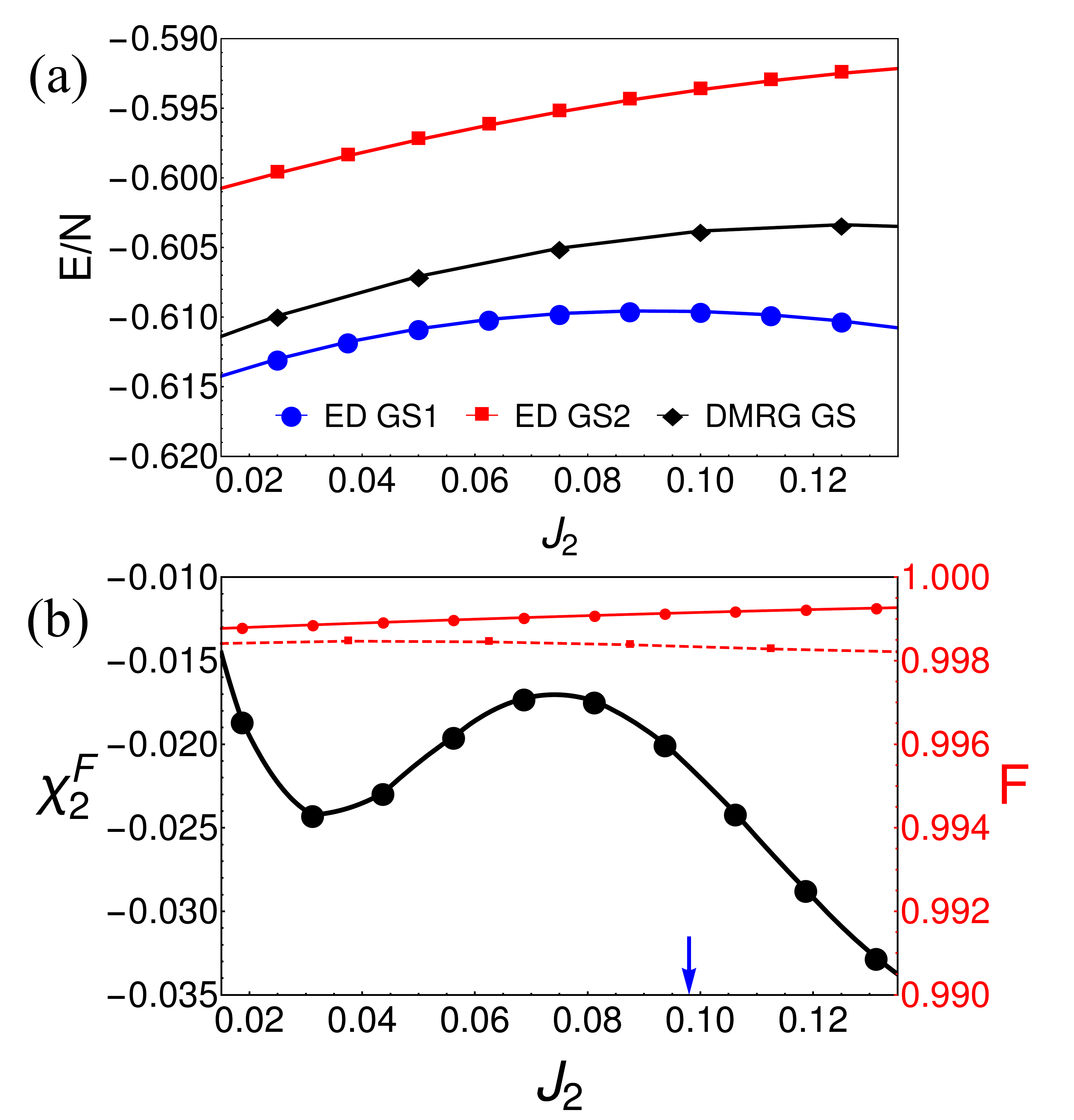}
\caption{(a) Energy per site for the triangular lattice model of a single ground state from DMRG and the two states from ED that make up the ground state manifold of the CSL. (b) Fidelity from DMRG (dashed red line) and ED (solid red line) and fidelity susceptibility from ED (solid black line). The transition is signified by a peak in the fidelity susceptibility at $J_2=0.075$. The arrow indicates the position at which the first excited state has a level crossing, switching from a singlet at the $\Gamma$ point (CSL) to a triplet at the $M$ point (tetrahedral). }
\label{fig:TriFid}
\end{figure}

At $J_2=0.15$, the system is in a tetrahedral state. Our ED results show that the ground state is a singlet while the first excited state is a spin triplet with 
momentum at the $M$ point. The static spin structure factor of the ground state, shown 
in Fig. \ref{fig:TriPD} (b), has clear peaks at the $M$ points. At the same time, DMRG finds a single ground state, indicative of a 
topologically trivial phase.

At $J_2=0.0$, the system is in the CSL phase. In ED, both the ground state and the first excited state are spin singlet states with momentum 
$\bk=(0,0)$. The static spin structure factor of the ground state, shown in Fig. \ref{fig:TriPD} (a), has no sharp peaks. Furthermore, in DMRG we find
two distinct ground states indicative of topological order, and the entanglement spectrum resembles that of a
free chiral boson, as shown in Fig. \ref{fig:TriPD} (c). Upon increasing $J_2$, and entering the tetrahedral state, we again find 
many additional low lying states (highlighted in Fig. \ref{fig:TriPD} (d)), leading to strong deviations from the free chiral boson description.

Again these results indicate that there must be a phase transition 
between the CSL and a magnetically ordered tetrahedral phase between $J_2=0.0$ and $J_2=0.15$. However, the energy of the ground states in ED and DMRG shown in 
Fig. \ref{fig:TriFid} show no sharp kinks and the fidelities are again smooth. It is only the fidelity susceptibility that provides a signal of the transition, with a broad peak 
centered at $J_2=0.075$. We thus again conclude that the CSL to tetrahedral transition is likely to be continuous, as indicated by our ED and DMRG signatures. 

We note that the signatures of the CSL, such as the structure of the entanglement spectrum, are not quite as clean here as in the honeycomb lattice case. This can be explained by the fact that the correlation length $\xi_\mathrm{TM}$, extracted from DMRG, is shorter for the honeycomb lattice case. The correlation length for the honeycomb lattice at $J_3=0.36$ is $\xi_\mathrm{TM}=0.88$ whereas for the triangular case at $J_2=0.0$ it is more than double at $\xi_\mathrm{TM}=1.89$. Here, $\xi_\mathrm{TM}$ denotes the transfer matrix correlation length defined as $\xi_\mathrm{TM} = -1/\log(\lambda_2)$, where $\lambda_2$ is the second largest eigenvalue of the transfer matrix. The transfer matrix contains tensors associated to one column of an infinite cylinder. $\xi_\mathrm{TM}$ is the upper bound for any correlation length in the system. The above values for $\xi_\mathrm{TM}$ are extrapolated in the bond dimension.

\subsection{Square Lattice}
\begin{figure}[tb]
\includegraphics[scale=0.25]{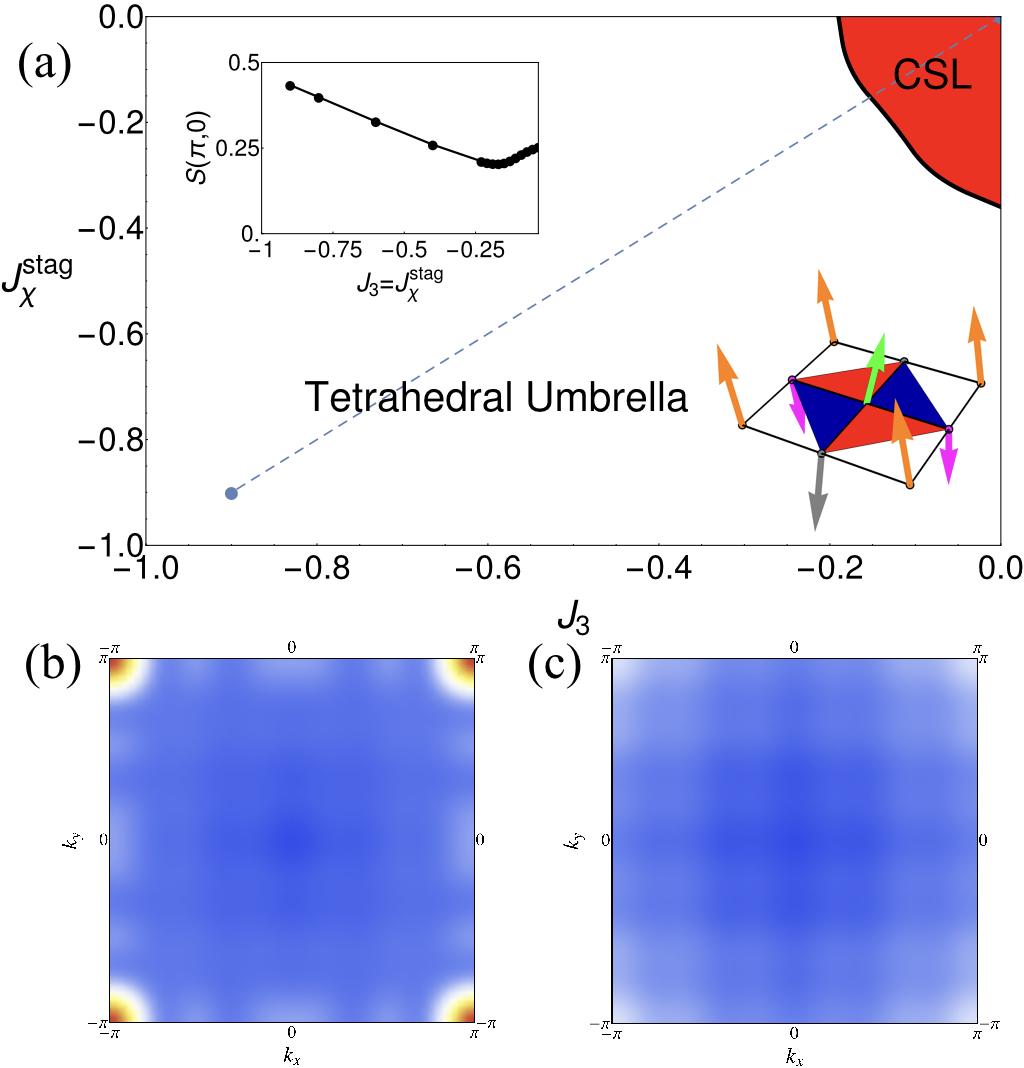}
\caption{(a) Phase diagram of the square lattice model for $N=36$ sites as a function of $J_3$ and $J_\chi^{\rm stag}$ 
for fixed $J_1=J_\chi=1.0$. The inset shows the evolution of the structure factor peak at $X=(\pi,0),(0,\pi)$ along the dashed line. The static spin structure 
factor of the ED ground state for $N=36$ spins is shown for the (b) tetrahedral and (c) CSL phases.}
\label{fig:SqPD}
\end{figure}

The square lattice $J_1,J_2$ model with a nonzero $J_\chi$ has already been shown to realize a CSL phase. \cite{Nielsen2013} Here, for simplicity, 
we set $J_1=J_\chi=1.0$, and $J_2=0$, which places us in the CSL phase. On the square lattice, there are no non-coplanar RMOs. However,
allowing for the breaking of $C_4$ symmetry about a lattice site allows for a non-coplanar RMO
which is called the `tetrahedral umbrella' state, depicted schematically in Fig.~\ref{fig:SqPD} (a). This state is a multimode spin crystal formed
by wavevectors at the $K=(\pi,\pi)$ and $X=(\pi,0),(0,\pi)$ points of the BZ. In addition to breaking spin rotational symmetry, the 
broken $C_4$ rotational symmetry about the square lattice sites leads to a staggered modulation of the scalar spin chirality. In order to
simply access this phase from the CSL, we add a ferromagnetic third neighbor Heisenberg term $J_3 < 0$
as well as a staggered chiral term $J_\chi^{\rm stag}$ that explicitly breaks the square lattice $C_4$ symmetry, 
which leads to the phase diagram depicted schematically in Fig.~\ref{fig:SqPD}(a).

\begin{figure}[tb]
\includegraphics[scale=0.2]{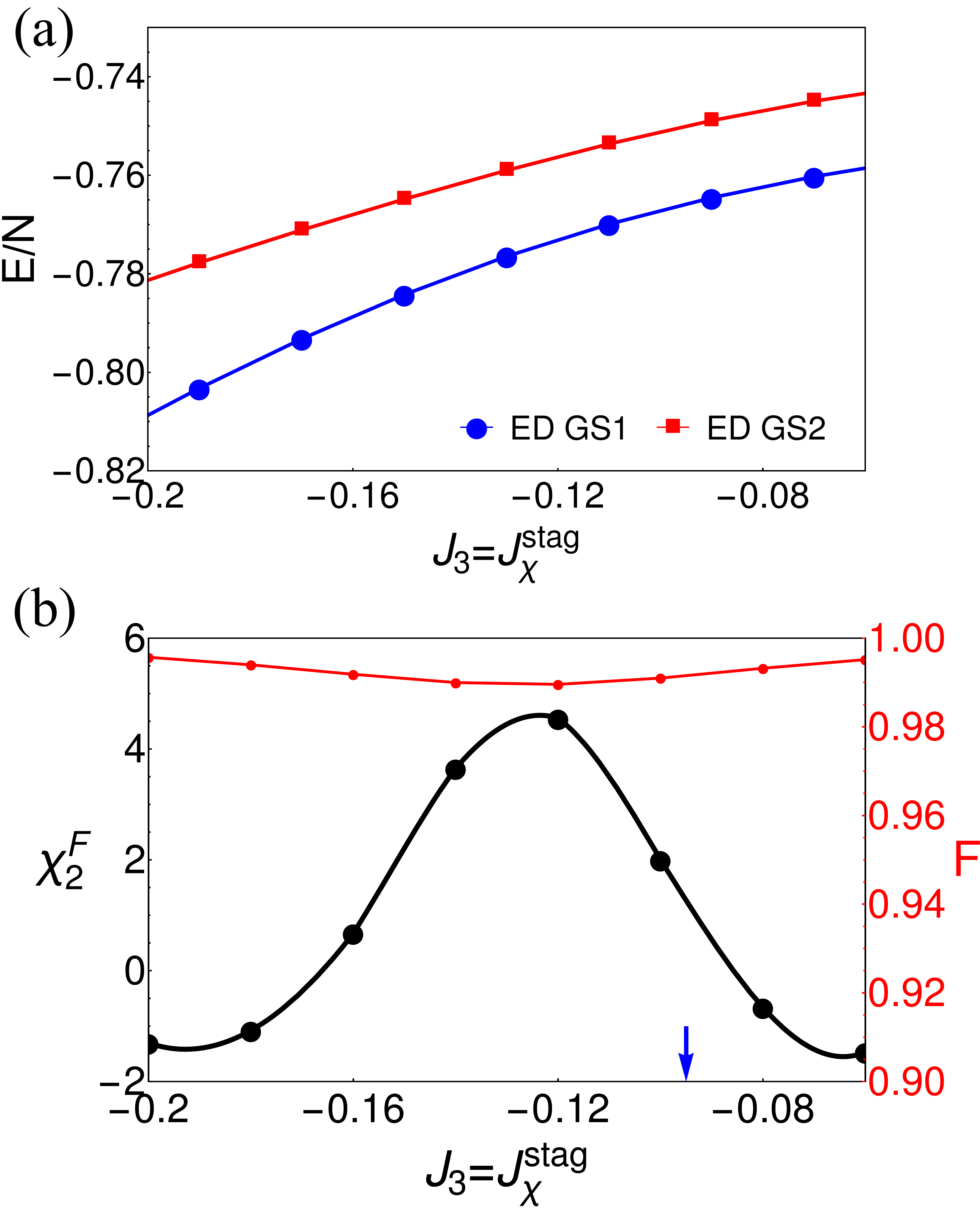}
\caption{(a) Energy of the two states from ED that make up the ground state manifold of the CSL for the square lattice model.
(b) Fidelity (solid red line) and fidelity susceptibility (solid black line) from ED with the peak in the fidelity susceptibility at $J_3=-0.125$ marking the transition from the CSL to the tetrahedral. The arrow indicates the position at which the first excited state switches from a singlet at the $\Gamma$ point to a triplet at the $K$-point. }
\label{fig:SqFid}
\end{figure}

For $J_3 = J_\chi^{\rm stag}=0$ the system is in the CSL phase. As before, the ED ground state and the first excited state are spin singlet states with momentum 
$\bk=(0,0)$, and they form the GSM of the CSL, which are reasonably well separated from all the other excited states. The static 
spin structure factor of the ground state, shown in Fig. \ref{fig:SqPD}(c), is featureless.
For significant $J_3 < 0$ and $J_\chi^{\rm stag}$, on the other hand, the system resembles the tetrahedral umbrella state. In ED, the ground state is a singlet
while the first excited state is now a spin triplet with momentum 
at the $K=(\pi,\pi)$ point, followed by slightly higher energy triplet excitations at the $X=(\pi,0),(0,\pi)$ points. The static spin structure factor of the ground state, shown 
in Fig. \ref{fig:SqPD}(b), has a sharp peak at the $K$ point and smaller peaks at the $X$ points. These suggest multimode order associated with the tetrahedral umbrella state.

In Fig. \ref{fig:SqFid}, we plot the energy per site which is smooth and the GSM fidelity from ED which has a weak dip indicating the transition, with
a clear peak in the fidelity susceptibility at $J_3=-0.125$. Just as in the honeycomb and triangular lattice, the CSL to tetrahedral transition does not show any 
signatures of first order behaviour, supporting the idea that it is again a continuous transition.

\section{Discussion}

We have provided numerical evidence (using ED and DMRG) that CSLs on the honeycomb, triangular, and square lattices arise from melting ordered non-coplanar spin 
crystal states via continuous transitions. This suggests that quantum melting such non-coplanar states is likely to be a general mechanism
to obtain CSLs.  In previous work, we have conjectured a possible Chern-Simons-Higgs theory which could possibly capture the tetrahedral-CSL transition on the honeycomb
lattice. Our work motivates a further study of this exotic continuous transition, including a comparison with the gauge theory spectrum on the torus which has been
studied within a large-${\cal N}$ approximation \cite{AlexTorus2016}.

There are a number of interesting extensions of this work to consider. In particular it is natural to ask whether this kind of mechanism could be extended to 3D 
lattices. Can quantum melting a 3D non-coplanar magnetic state lead to a stacked chiral spin liquid state? This may be of relevance to the 3D pyrochlore 
Pr$_2$Ir$_2$O$_7$, which appears to exhibit an anomalous Hall effect without long-range 
magnetic order, suggestive perhaps of a spin liquid state with nonzero scalar spin chirality.\cite{Nakatsuji2010} 
Another interesting direction would be to explore higher spin systems. 
On the ordered 
side, moving to higher spin allows for a richer set of order parameters, such as quadrupolar order in the case of spin-$1$. On the CSL side, higher spin CSLs can 
have non-Abelian topological order and thus a more complex GSM structure. \cite{GreiterNACSL2009,GreiterNACSL2011,GreiterNACSL2014,TKNg2015,Tsvelik2017} Is it possible to find an ordered ``parent state" for these higher spin non-Abelian CSLs?

\textit{Note Added} - During completion of this work, we became aware of two recent preprints \cite{McCulloch2017,ShengDMRG2017} that also discuss the $J_1$-$J_2$-$J_\chi$ model on the 
triangular lattice. Our results are consistent where there is overlap. 

\section{Acknowledgements}

We thank L. Balents, I. Martin, and S. Bhattacharjee for useful discussions. CH and AP acknowledge support from NSERC of Canada. LC acknowledges support 
by the John Templeton Foundation. Z.P. acknowledges support by EPSRC grant EP/P009409/1. Statement of compliance with EPSRC policy framework 
on research data: This publication is theoretical work that does not require supporting research data. This research was 
supported in part by Perimeter Institute for Theoretical Physics. Research at Perimeter Institute is supported 
by the Government of Canada through Industry Canada and by the Province of Ontario through the Ministry of Research and Innovation. This research is supported in part by the U.S. Department of Energy through J. Robert Oppenheimer fellowship (LC). Computations were performed on the GPC supercomputer at the 
SciNet HPC Consortium. SciNet is funded by: the Canada Foundation for Innovation under the auspices of Compute Canada; the Government of Ontario; Ontario 
Research Fund - Research Excellence; and the University of Toronto. This work also made use of the facilities of N8 HPC Centre of Excellence, provided and funded 
by the N8 consortium and EPSRC (Grant No.EP/K000225/1). The Centre is co-ordinated by the Universities of Leeds and Manchester. Some of the computations were done on the
supercomputer Mammouth parall\`ele II from University
of Sherbrooke, managed by Calcul Qu\'ebec and Compute
Canada. The operation of this supercomputer is funded
by the Canada Foundation for Innovation (CFI), the minist\`ere
de l'\'Economie, de la science et de l'innovation du Qu\'ebec (MESI) and the Fonds de recherche du Qu\'ebec -
Nature et technologies (FRQ-NT).


\begin{thebibliography}{56}%
\makeatletter
\providecommand \@ifxundefined [1]{%
 \@ifx{#1\undefined}
}%
\providecommand \@ifnum [1]{%
 \ifnum #1\expandafter \@firstoftwo
 \else \expandafter \@secondoftwo
 \fi
}%
\providecommand \@ifx [1]{%
 \ifx #1\expandafter \@firstoftwo
 \else \expandafter \@secondoftwo
 \fi
}%
\providecommand \natexlab [1]{#1}%
\providecommand \enquote  [1]{``#1''}%
\providecommand \bibnamefont  [1]{#1}%
\providecommand \bibfnamefont [1]{#1}%
\providecommand \citenamefont [1]{#1}%
\providecommand \href@noop [0]{\@secondoftwo}%
\providecommand \href [0]{\begingroup \@sanitize@url \@href}%
\providecommand \@href[1]{\@@startlink{#1}\@@href}%
\providecommand \@@href[1]{\endgroup#1\@@endlink}%
\providecommand \@sanitize@url [0]{\catcode `\\12\catcode `\$12\catcode
  `\&12\catcode `\#12\catcode `\^12\catcode `\_12\catcode `\%12\relax}%
\providecommand \@@startlink[1]{}%
\providecommand \@@endlink[0]{}%
\providecommand \url  [0]{\begingroup\@sanitize@url \@url }%
\providecommand \@url [1]{\endgroup\@href {#1}{\urlprefix }}%
\providecommand \urlprefix  [0]{URL }%
\providecommand \Eprint [0]{\href }%
\providecommand \doibase [0]{http://dx.doi.org/}%
\providecommand \selectlanguage [0]{\@gobble}%
\providecommand \bibinfo  [0]{\@secondoftwo}%
\providecommand \bibfield  [0]{\@secondoftwo}%
\providecommand \translation [1]{[#1]}%
\providecommand \BibitemOpen [0]{}%
\providecommand \bibitemStop [0]{}%
\providecommand \bibitemNoStop [0]{.\EOS\space}%
\providecommand \EOS [0]{\spacefactor3000\relax}%
\providecommand \BibitemShut  [1]{\csname bibitem#1\endcsname}%
\let\auto@bib@innerbib\@empty
\bibitem [{\citenamefont {Zhou}\ \emph {et~al.}(2017)\citenamefont {Zhou},
  \citenamefont {Kanoda},\ and\ \citenamefont {Ng}}]{QSLRMP2017}%
  \BibitemOpen
  \bibfield  {author} {\bibinfo {author} {\bibfnamefont {Y.}~\bibnamefont
  {Zhou}}, \bibinfo {author} {\bibfnamefont {K.}~\bibnamefont {Kanoda}}, \ and\
  \bibinfo {author} {\bibfnamefont {T.-K.}\ \bibnamefont {Ng}},\ }\href
  {\doibase 10.1103/RevModPhys.89.025003} {\bibfield  {journal} {\bibinfo
  {journal} {Rev. Mod. Phys.}\ }\textbf {\bibinfo {volume} {89}},\ \bibinfo
  {pages} {025003} (\bibinfo {year} {2017})}\BibitemShut {NoStop}%
\bibitem [{\citenamefont {Savary}\ and\ \citenamefont
  {Balents}(2017)}]{Savary2017}%
  \BibitemOpen
  \bibfield  {author} {\bibinfo {author} {\bibfnamefont {L.}~\bibnamefont
  {Savary}}\ and\ \bibinfo {author} {\bibfnamefont {L.}~\bibnamefont
  {Balents}},\ }\href {http://stacks.iop.org/0034-4885/80/i=1/a=016502}
  {\bibfield  {journal} {\bibinfo  {journal} {Reports on Progress in Physics}\
  }\textbf {\bibinfo {volume} {80}},\ \bibinfo {pages} {016502} (\bibinfo
  {year} {2017})}\BibitemShut {NoStop}%
\bibitem [{\citenamefont {Nayak}\ \emph {et~al.}(2008)\citenamefont {Nayak},
  \citenamefont {Simon}, \citenamefont {Stern}, \citenamefont {Freedman},\ and\
  \citenamefont {Das~Sarma}}]{NayakRMP2008}%
  \BibitemOpen
  \bibfield  {author} {\bibinfo {author} {\bibfnamefont {C.}~\bibnamefont
  {Nayak}}, \bibinfo {author} {\bibfnamefont {S.~H.}\ \bibnamefont {Simon}},
  \bibinfo {author} {\bibfnamefont {A.}~\bibnamefont {Stern}}, \bibinfo
  {author} {\bibfnamefont {M.}~\bibnamefont {Freedman}}, \ and\ \bibinfo
  {author} {\bibfnamefont {S.}~\bibnamefont {Das~Sarma}},\ }\href {\doibase
  10.1103/RevModPhys.80.1083} {\bibfield  {journal} {\bibinfo  {journal} {Rev.
  Mod. Phys.}\ }\textbf {\bibinfo {volume} {80}},\ \bibinfo {pages} {1083}
  (\bibinfo {year} {2008})}\BibitemShut {NoStop}%
\bibitem [{\citenamefont {Kalmeyer}\ and\ \citenamefont
  {Laughlin}(1987)}]{KalmeyerLaughlin}%
  \BibitemOpen
  \bibfield  {author} {\bibinfo {author} {\bibfnamefont {V.}~\bibnamefont
  {Kalmeyer}}\ and\ \bibinfo {author} {\bibfnamefont {R.~B.}\ \bibnamefont
  {Laughlin}},\ }\href {\doibase 10.1103/PhysRevLett.59.2095} {\bibfield
  {journal} {\bibinfo  {journal} {Phys. Rev. Lett.}\ }\textbf {\bibinfo
  {volume} {59}},\ \bibinfo {pages} {2095} (\bibinfo {year}
  {1987})}\BibitemShut {NoStop}%
\bibitem [{\citenamefont {Schroeter}\ \emph {et~al.}(2007)\citenamefont
  {Schroeter}, \citenamefont {Kapit}, \citenamefont {Thomale},\ and\
  \citenamefont {Greiter}}]{SchroeterCSL2007}%
  \BibitemOpen
  \bibfield  {author} {\bibinfo {author} {\bibfnamefont {D.~F.}\ \bibnamefont
  {Schroeter}}, \bibinfo {author} {\bibfnamefont {E.}~\bibnamefont {Kapit}},
  \bibinfo {author} {\bibfnamefont {R.}~\bibnamefont {Thomale}}, \ and\
  \bibinfo {author} {\bibfnamefont {M.}~\bibnamefont {Greiter}},\ }\href
  {\doibase 10.1103/PhysRevLett.99.097202} {\bibfield  {journal} {\bibinfo
  {journal} {Phys. Rev. Lett.}\ }\textbf {\bibinfo {volume} {99}},\ \bibinfo
  {pages} {097202} (\bibinfo {year} {2007})}\BibitemShut {NoStop}%
\bibitem [{\citenamefont {Thomale}\ \emph {et~al.}(2009)\citenamefont
  {Thomale}, \citenamefont {Kapit}, \citenamefont {Schroeter},\ and\
  \citenamefont {Greiter}}]{ThomaleCSL2009}%
  \BibitemOpen
  \bibfield  {author} {\bibinfo {author} {\bibfnamefont {R.}~\bibnamefont
  {Thomale}}, \bibinfo {author} {\bibfnamefont {E.}~\bibnamefont {Kapit}},
  \bibinfo {author} {\bibfnamefont {D.~F.}\ \bibnamefont {Schroeter}}, \ and\
  \bibinfo {author} {\bibfnamefont {M.}~\bibnamefont {Greiter}},\ }\href
  {\doibase 10.1103/PhysRevB.80.104406} {\bibfield  {journal} {\bibinfo
  {journal} {Phys. Rev. B}\ }\textbf {\bibinfo {volume} {80}},\ \bibinfo
  {pages} {104406} (\bibinfo {year} {2009})}\BibitemShut {NoStop}%
\bibitem [{\citenamefont {Greiter}\ \emph
  {et~al.}(2014{\natexlab{a}})\citenamefont {Greiter}, \citenamefont
  {Schroeter},\ and\ \citenamefont {Thomale}}]{ThomaleNACSL2014}%
  \BibitemOpen
  \bibfield  {author} {\bibinfo {author} {\bibfnamefont {M.}~\bibnamefont
  {Greiter}}, \bibinfo {author} {\bibfnamefont {D.~F.}\ \bibnamefont
  {Schroeter}}, \ and\ \bibinfo {author} {\bibfnamefont {R.}~\bibnamefont
  {Thomale}},\ }\href {\doibase 10.1103/PhysRevB.89.165125} {\bibfield
  {journal} {\bibinfo  {journal} {Phys. Rev. B}\ }\textbf {\bibinfo {volume}
  {89}},\ \bibinfo {pages} {165125} (\bibinfo {year}
  {2014}{\natexlab{a}})}\BibitemShut {NoStop}%
\bibitem [{\citenamefont {Meng}\ \emph {et~al.}(2015)\citenamefont {Meng},
  \citenamefont {Neupert}, \citenamefont {Greiter},\ and\ \citenamefont
  {Thomale}}]{Meng2015}%
  \BibitemOpen
  \bibfield  {author} {\bibinfo {author} {\bibfnamefont {T.}~\bibnamefont
  {Meng}}, \bibinfo {author} {\bibfnamefont {T.}~\bibnamefont {Neupert}},
  \bibinfo {author} {\bibfnamefont {M.}~\bibnamefont {Greiter}}, \ and\
  \bibinfo {author} {\bibfnamefont {R.}~\bibnamefont {Thomale}},\ }\href
  {\doibase 10.1103/PhysRevB.91.241106} {\bibfield  {journal} {\bibinfo
  {journal} {Phys. Rev. B}\ }\textbf {\bibinfo {volume} {91}},\ \bibinfo
  {pages} {241106} (\bibinfo {year} {2015})}\BibitemShut {NoStop}%
\bibitem [{\citenamefont {He}\ \emph {et~al.}(2014)\citenamefont {He},
  \citenamefont {Sheng},\ and\ \citenamefont {Chen}}]{YHe2014}%
  \BibitemOpen
  \bibfield  {author} {\bibinfo {author} {\bibfnamefont {Y.-C.}\ \bibnamefont
  {He}}, \bibinfo {author} {\bibfnamefont {D.~N.}\ \bibnamefont {Sheng}}, \
  and\ \bibinfo {author} {\bibfnamefont {Y.}~\bibnamefont {Chen}},\ }\href
  {\doibase 10.1103/PhysRevLett.112.137202} {\bibfield  {journal} {\bibinfo
  {journal} {Phys. Rev. Lett.}\ }\textbf {\bibinfo {volume} {112}},\ \bibinfo
  {pages} {137202} (\bibinfo {year} {2014})}\BibitemShut {NoStop}%
\bibitem [{\citenamefont {Bauer}\ \emph {et~al.}(2014)\citenamefont {Bauer},
  \citenamefont {Cincio}, \citenamefont {Keller}, \citenamefont {Dolfi},
  \citenamefont {Vidal}, \citenamefont {Trebst},\ and\ \citenamefont
  {Ludwig}}]{Bauer2014}%
  \BibitemOpen
  \bibfield  {author} {\bibinfo {author} {\bibfnamefont {B.}~\bibnamefont
  {Bauer}}, \bibinfo {author} {\bibfnamefont {L.}~\bibnamefont {Cincio}},
  \bibinfo {author} {\bibfnamefont {B.~P.}\ \bibnamefont {Keller}}, \bibinfo
  {author} {\bibfnamefont {M.}~\bibnamefont {Dolfi}}, \bibinfo {author}
  {\bibfnamefont {G.}~\bibnamefont {Vidal}}, \bibinfo {author} {\bibfnamefont
  {S.}~\bibnamefont {Trebst}}, \ and\ \bibinfo {author} {\bibfnamefont
  {A.~W.~W.}\ \bibnamefont {Ludwig}},\ }\href
  {http://dx.doi.org/10.1038/ncomms6137} {\bibfield  {journal} {\bibinfo
  {journal} {Nat Commun}\ }\textbf {\bibinfo {volume} {5}},\ \bibinfo {pages}
  {5137} (\bibinfo {year} {2014})}\BibitemShut {NoStop}%
\bibitem [{\citenamefont {Gong}\ \emph {et~al.}(2014)\citenamefont {Gong},
  \citenamefont {Zhu},\ and\ \citenamefont {Sheng}}]{Gong2014}%
  \BibitemOpen
  \bibfield  {author} {\bibinfo {author} {\bibfnamefont {S.-S.}\ \bibnamefont
  {Gong}}, \bibinfo {author} {\bibfnamefont {W.}~\bibnamefont {Zhu}}, \ and\
  \bibinfo {author} {\bibfnamefont {D.~N.}\ \bibnamefont {Sheng}},\ }\href
  {http://dx.doi.org/10.1038/srep06317} {\bibfield  {journal} {\bibinfo
  {journal} {Scientific Reports}\ }\textbf {\bibinfo {volume} {4}},\ \bibinfo
  {pages} {6317} (\bibinfo {year} {2014})}\BibitemShut {NoStop}%
\bibitem [{\citenamefont {Kumar}\ \emph {et~al.}(2014)\citenamefont {Kumar},
  \citenamefont {Sun},\ and\ \citenamefont {Fradkin}}]{FradkinCSL2014}%
  \BibitemOpen
  \bibfield  {author} {\bibinfo {author} {\bibfnamefont {K.}~\bibnamefont
  {Kumar}}, \bibinfo {author} {\bibfnamefont {K.}~\bibnamefont {Sun}}, \ and\
  \bibinfo {author} {\bibfnamefont {E.}~\bibnamefont {Fradkin}},\ }\href
  {\doibase 10.1103/PhysRevB.90.174409} {\bibfield  {journal} {\bibinfo
  {journal} {Phys. Rev. B}\ }\textbf {\bibinfo {volume} {90}},\ \bibinfo
  {pages} {174409} (\bibinfo {year} {2014})}\BibitemShut {NoStop}%
\bibitem [{\citenamefont {Gong}\ \emph {et~al.}(2015)\citenamefont {Gong},
  \citenamefont {Zhu}, \citenamefont {Balents},\ and\ \citenamefont
  {Sheng}}]{Gong2015}%
  \BibitemOpen
  \bibfield  {author} {\bibinfo {author} {\bibfnamefont {S.-S.}\ \bibnamefont
  {Gong}}, \bibinfo {author} {\bibfnamefont {W.}~\bibnamefont {Zhu}}, \bibinfo
  {author} {\bibfnamefont {L.}~\bibnamefont {Balents}}, \ and\ \bibinfo
  {author} {\bibfnamefont {D.~N.}\ \bibnamefont {Sheng}},\ }\href {\doibase
  10.1103/PhysRevB.91.075112} {\bibfield  {journal} {\bibinfo  {journal} {Phys.
  Rev. B}\ }\textbf {\bibinfo {volume} {91}},\ \bibinfo {pages} {075112}
  (\bibinfo {year} {2015})}\BibitemShut {NoStop}%
\bibitem [{\citenamefont {Wietek}\ \emph {et~al.}(2015)\citenamefont {Wietek},
  \citenamefont {Sterdyniak},\ and\ \citenamefont {L\"auchli}}]{Wietek2015}%
  \BibitemOpen
  \bibfield  {author} {\bibinfo {author} {\bibfnamefont {A.}~\bibnamefont
  {Wietek}}, \bibinfo {author} {\bibfnamefont {A.}~\bibnamefont {Sterdyniak}},
  \ and\ \bibinfo {author} {\bibfnamefont {A.~M.}\ \bibnamefont {L\"auchli}},\
  }\href {\doibase 10.1103/PhysRevB.92.125122} {\bibfield  {journal} {\bibinfo
  {journal} {Phys. Rev. B}\ }\textbf {\bibinfo {volume} {92}},\ \bibinfo
  {pages} {125122} (\bibinfo {year} {2015})}\BibitemShut {NoStop}%
\bibitem [{\citenamefont {Zhu}\ \emph {et~al.}(2016)\citenamefont {Zhu},
  \citenamefont {Gong},\ and\ \citenamefont {Sheng}}]{DNShengKagome2015}%
  \BibitemOpen
  \bibfield  {author} {\bibinfo {author} {\bibfnamefont {W.}~\bibnamefont
  {Zhu}}, \bibinfo {author} {\bibfnamefont {S.~S.}\ \bibnamefont {Gong}}, \
  and\ \bibinfo {author} {\bibfnamefont {D.~N.}\ \bibnamefont {Sheng}},\ }\href
  {\doibase 10.1103/PhysRevB.94.035129} {\bibfield  {journal} {\bibinfo
  {journal} {Phys. Rev. B}\ }\textbf {\bibinfo {volume} {94}},\ \bibinfo
  {pages} {035129} (\bibinfo {year} {2016})}\BibitemShut {NoStop}%
\bibitem [{\citenamefont {Hu}\ \emph {et~al.}(2015)\citenamefont {Hu},
  \citenamefont {Zhu}, \citenamefont {Zhang}, \citenamefont {Gong},
  \citenamefont {Becca},\ and\ \citenamefont {Sheng}}]{ShengVMC2015}%
  \BibitemOpen
  \bibfield  {author} {\bibinfo {author} {\bibfnamefont {W.-J.}\ \bibnamefont
  {Hu}}, \bibinfo {author} {\bibfnamefont {W.}~\bibnamefont {Zhu}}, \bibinfo
  {author} {\bibfnamefont {Y.}~\bibnamefont {Zhang}}, \bibinfo {author}
  {\bibfnamefont {S.}~\bibnamefont {Gong}}, \bibinfo {author} {\bibfnamefont
  {F.}~\bibnamefont {Becca}}, \ and\ \bibinfo {author} {\bibfnamefont {D.~N.}\
  \bibnamefont {Sheng}},\ }\href {\doibase 10.1103/PhysRevB.91.041124}
  {\bibfield  {journal} {\bibinfo  {journal} {Phys. Rev. B}\ }\textbf {\bibinfo
  {volume} {91}},\ \bibinfo {pages} {041124} (\bibinfo {year}
  {2015})}\BibitemShut {NoStop}%
\bibitem [{\citenamefont {Bieri}\ \emph {et~al.}(2015)\citenamefont {Bieri},
  \citenamefont {Messio}, \citenamefont {Bernu},\ and\ \citenamefont
  {Lhuillier}}]{BieriVWFKag2015}%
  \BibitemOpen
  \bibfield  {author} {\bibinfo {author} {\bibfnamefont {S.}~\bibnamefont
  {Bieri}}, \bibinfo {author} {\bibfnamefont {L.}~\bibnamefont {Messio}},
  \bibinfo {author} {\bibfnamefont {B.}~\bibnamefont {Bernu}}, \ and\ \bibinfo
  {author} {\bibfnamefont {C.}~\bibnamefont {Lhuillier}},\ }\href {\doibase
  10.1103/PhysRevB.92.060407} {\bibfield  {journal} {\bibinfo  {journal} {Phys.
  Rev. B}\ }\textbf {\bibinfo {volume} {92}},\ \bibinfo {pages} {060407}
  (\bibinfo {year} {2015})}\BibitemShut {NoStop}%
\bibitem [{\citenamefont {Kumar}\ \emph {et~al.}(2015)\citenamefont {Kumar},
  \citenamefont {Sun},\ and\ \citenamefont {Fradkin}}]{FradkinCSL2015}%
  \BibitemOpen
  \bibfield  {author} {\bibinfo {author} {\bibfnamefont {K.}~\bibnamefont
  {Kumar}}, \bibinfo {author} {\bibfnamefont {K.}~\bibnamefont {Sun}}, \ and\
  \bibinfo {author} {\bibfnamefont {E.}~\bibnamefont {Fradkin}},\ }\href
  {\doibase 10.1103/PhysRevB.92.094433} {\bibfield  {journal} {\bibinfo
  {journal} {Phys. Rev. B}\ }\textbf {\bibinfo {volume} {92}},\ \bibinfo
  {pages} {094433} (\bibinfo {year} {2015})}\BibitemShut {NoStop}%
\bibitem [{\citenamefont {Nielsen}\ \emph {et~al.}(2013)\citenamefont
  {Nielsen}, \citenamefont {Sierra},\ and\ \citenamefont
  {Cirac}}]{Nielsen2013}%
  \BibitemOpen
  \bibfield  {author} {\bibinfo {author} {\bibfnamefont {A.~E.~B.}\
  \bibnamefont {Nielsen}}, \bibinfo {author} {\bibfnamefont {G.}~\bibnamefont
  {Sierra}}, \ and\ \bibinfo {author} {\bibfnamefont {J.~I.}\ \bibnamefont
  {Cirac}},\ }\href {http://dx.doi.org/10.1038/ncomms3864} {\bibfield
  {journal} {\bibinfo  {journal} {Nat Commun}\ }\textbf {\bibinfo {volume} {4}}
  (\bibinfo {year} {2013})}\BibitemShut {NoStop}%
\bibitem [{\citenamefont {Poilblanc}\ \emph {et~al.}(2015)\citenamefont
  {Poilblanc}, \citenamefont {Cirac},\ and\ \citenamefont
  {Schuch}}]{Poilblanc2015}%
  \BibitemOpen
  \bibfield  {author} {\bibinfo {author} {\bibfnamefont {D.}~\bibnamefont
  {Poilblanc}}, \bibinfo {author} {\bibfnamefont {J.~I.}\ \bibnamefont
  {Cirac}}, \ and\ \bibinfo {author} {\bibfnamefont {N.}~\bibnamefont
  {Schuch}},\ }\href {\doibase 10.1103/PhysRevB.91.224431} {\bibfield
  {journal} {\bibinfo  {journal} {Phys. Rev. B}\ }\textbf {\bibinfo {volume}
  {91}},\ \bibinfo {pages} {224431} (\bibinfo {year} {2015})}\BibitemShut
  {NoStop}%
\bibitem [{\citenamefont {Liu}\ \emph {et~al.}(2016)\citenamefont {Liu},
  \citenamefont {Liu}, \citenamefont {Law}, \citenamefont {Liu},\ and\
  \citenamefont {Ng}}]{XJLiu2016}%
  \BibitemOpen
  \bibfield  {author} {\bibinfo {author} {\bibfnamefont {X.-J.}\ \bibnamefont
  {Liu}}, \bibinfo {author} {\bibfnamefont {Z.-X.}\ \bibnamefont {Liu}},
  \bibinfo {author} {\bibfnamefont {K.~T.}\ \bibnamefont {Law}}, \bibinfo
  {author} {\bibfnamefont {W.~V.}\ \bibnamefont {Liu}}, \ and\ \bibinfo
  {author} {\bibfnamefont {T.~K.}\ \bibnamefont {Ng}},\ }\href
  {http://stacks.iop.org/1367-2630/18/i=3/a=035004} {\bibfield  {journal}
  {\bibinfo  {journal} {New Journal of Physics}\ }\textbf {\bibinfo {volume}
  {18}},\ \bibinfo {pages} {035004} (\bibinfo {year} {2016})}\BibitemShut
  {NoStop}%
\bibitem [{\citenamefont {Hickey}\ \emph {et~al.}(2016)\citenamefont {Hickey},
  \citenamefont {Cincio}, \citenamefont {Papi\ifmmode~\acute{c}\else
  \'{c}\fi{}},\ and\ \citenamefont {Paramekanti}}]{Hickey2016}%
  \BibitemOpen
  \bibfield  {author} {\bibinfo {author} {\bibfnamefont {C.}~\bibnamefont
  {Hickey}}, \bibinfo {author} {\bibfnamefont {L.}~\bibnamefont {Cincio}},
  \bibinfo {author} {\bibfnamefont {Z.}~\bibnamefont
  {Papi\ifmmode~\acute{c}\else \'{c}\fi{}}}, \ and\ \bibinfo {author}
  {\bibfnamefont {A.}~\bibnamefont {Paramekanti}},\ }\href {\doibase
  10.1103/PhysRevLett.116.137202} {\bibfield  {journal} {\bibinfo  {journal}
  {Phys. Rev. Lett.}\ }\textbf {\bibinfo {volume} {116}},\ \bibinfo {pages}
  {137202} (\bibinfo {year} {2016})}\BibitemShut {NoStop}%
\bibitem [{\citenamefont {{Hu}}\ \emph {et~al.}(2016)\citenamefont {{Hu}},
  \citenamefont {{Gong}},\ and\ \citenamefont {{Sheng}}}]{ShengVMC2016}%
  \BibitemOpen
  \bibfield  {author} {\bibinfo {author} {\bibfnamefont {W.-J.}\ \bibnamefont
  {{Hu}}}, \bibinfo {author} {\bibfnamefont {S.-S.}\ \bibnamefont {{Gong}}}, \
  and\ \bibinfo {author} {\bibfnamefont {D.~N.}\ \bibnamefont {{Sheng}}},\
  }\href@noop {} {\bibfield  {journal} {\bibinfo  {journal} {ArXiv e-prints}\ }
  (\bibinfo {year} {2016})},\ \Eprint {http://arxiv.org/abs/1603.03365}
  {arXiv:1603.03365 [cond-mat.str-el]} \BibitemShut {NoStop}%
\bibitem [{\citenamefont {Wietek}\ and\ \citenamefont
  {L\"auchli}(2017)}]{Wietek2016}%
  \BibitemOpen
  \bibfield  {author} {\bibinfo {author} {\bibfnamefont {A.}~\bibnamefont
  {Wietek}}\ and\ \bibinfo {author} {\bibfnamefont {A.~M.}\ \bibnamefont
  {L\"auchli}},\ }\href {\doibase 10.1103/PhysRevB.95.035141} {\bibfield
  {journal} {\bibinfo  {journal} {Phys. Rev. B}\ }\textbf {\bibinfo {volume}
  {95}},\ \bibinfo {pages} {035141} (\bibinfo {year} {2017})}\BibitemShut
  {NoStop}%
\bibitem [{\citenamefont {Wen}\ \emph {et~al.}(1989)\citenamefont {Wen},
  \citenamefont {Wilczek},\ and\ \citenamefont {Zee}}]{WenCSL1989}%
  \BibitemOpen
  \bibfield  {author} {\bibinfo {author} {\bibfnamefont {X.~G.}\ \bibnamefont
  {Wen}}, \bibinfo {author} {\bibfnamefont {F.}~\bibnamefont {Wilczek}}, \ and\
  \bibinfo {author} {\bibfnamefont {A.}~\bibnamefont {Zee}},\ }\href {\doibase
  10.1103/PhysRevB.39.11413} {\bibfield  {journal} {\bibinfo  {journal} {Phys.
  Rev. B}\ }\textbf {\bibinfo {volume} {39}},\ \bibinfo {pages} {11413}
  (\bibinfo {year} {1989})}\BibitemShut {NoStop}%
\bibitem [{\citenamefont {Wen}(1991)}]{WenCSL1991}%
  \BibitemOpen
  \bibfield  {author} {\bibinfo {author} {\bibfnamefont {X.~G.}\ \bibnamefont
  {Wen}},\ }\href {\doibase 10.1103/PhysRevB.44.2664} {\bibfield  {journal}
  {\bibinfo  {journal} {Phys. Rev. B}\ }\textbf {\bibinfo {volume} {44}},\
  \bibinfo {pages} {2664} (\bibinfo {year} {1991})}\BibitemShut {NoStop}%
\bibitem [{\citenamefont {Zhang}\ \emph {et~al.}(2011)\citenamefont {Zhang},
  \citenamefont {Grover},\ and\ \citenamefont {Vishwanath}}]{Zhang2011}%
  \BibitemOpen
  \bibfield  {author} {\bibinfo {author} {\bibfnamefont {Y.}~\bibnamefont
  {Zhang}}, \bibinfo {author} {\bibfnamefont {T.}~\bibnamefont {Grover}}, \
  and\ \bibinfo {author} {\bibfnamefont {A.}~\bibnamefont {Vishwanath}},\
  }\href {\doibase 10.1103/PhysRevB.84.075128} {\bibfield  {journal} {\bibinfo
  {journal} {Phys. Rev. B}\ }\textbf {\bibinfo {volume} {84}},\ \bibinfo
  {pages} {075128} (\bibinfo {year} {2011})}\BibitemShut {NoStop}%
\bibitem [{\citenamefont {Zhang}\ \emph {et~al.}(2012)\citenamefont {Zhang},
  \citenamefont {Grover}, \citenamefont {Turner}, \citenamefont {Oshikawa},\
  and\ \citenamefont {Vishwanath}}]{Zhang2012}%
  \BibitemOpen
  \bibfield  {author} {\bibinfo {author} {\bibfnamefont {Y.}~\bibnamefont
  {Zhang}}, \bibinfo {author} {\bibfnamefont {T.}~\bibnamefont {Grover}},
  \bibinfo {author} {\bibfnamefont {A.}~\bibnamefont {Turner}}, \bibinfo
  {author} {\bibfnamefont {M.}~\bibnamefont {Oshikawa}}, \ and\ \bibinfo
  {author} {\bibfnamefont {A.}~\bibnamefont {Vishwanath}},\ }\href {\doibase
  10.1103/PhysRevB.85.235151} {\bibfield  {journal} {\bibinfo  {journal} {Phys.
  Rev. B}\ }\textbf {\bibinfo {volume} {85}},\ \bibinfo {pages} {235151}
  (\bibinfo {year} {2012})}\BibitemShut {NoStop}%
\bibitem [{\citenamefont {{Barkeshli}}(2013)}]{Barkeshli2013}%
  \BibitemOpen
  \bibfield  {author} {\bibinfo {author} {\bibfnamefont {M.}~\bibnamefont
  {{Barkeshli}}},\ }\href@noop {} {\bibfield  {journal} {\bibinfo  {journal}
  {ArXiv e-prints}\ } (\bibinfo {year} {2013})},\ \Eprint
  {http://arxiv.org/abs/1307.8194} {arXiv:1307.8194 [cond-mat.str-el]}
  \BibitemShut {NoStop}%
\bibitem [{\citenamefont {He}\ \emph {et~al.}(2015)\citenamefont {He},
  \citenamefont {Bhattacharjee}, \citenamefont {Pollmann},\ and\ \citenamefont
  {Moessner}}]{YCHe2015}%
  \BibitemOpen
  \bibfield  {author} {\bibinfo {author} {\bibfnamefont {Y.-C.}\ \bibnamefont
  {He}}, \bibinfo {author} {\bibfnamefont {S.}~\bibnamefont {Bhattacharjee}},
  \bibinfo {author} {\bibfnamefont {F.}~\bibnamefont {Pollmann}}, \ and\
  \bibinfo {author} {\bibfnamefont {R.}~\bibnamefont {Moessner}},\ }\href
  {\doibase 10.1103/PhysRevLett.115.267209} {\bibfield  {journal} {\bibinfo
  {journal} {Phys. Rev. Lett.}\ }\textbf {\bibinfo {volume} {115}},\ \bibinfo
  {pages} {267209} (\bibinfo {year} {2015})}\BibitemShut {NoStop}%
\bibitem [{\citenamefont {Moessner}\ and\ \citenamefont
  {Sondhi}(2001)}]{Moessner01}%
  \BibitemOpen
  \bibfield  {author} {\bibinfo {author} {\bibfnamefont {R.}~\bibnamefont
  {Moessner}}\ and\ \bibinfo {author} {\bibfnamefont {S.~L.}\ \bibnamefont
  {Sondhi}},\ }\href {\doibase 10.1103/PhysRevLett.86.1881} {\bibfield
  {journal} {\bibinfo  {journal} {Phys. Rev. Lett.}\ }\textbf {\bibinfo
  {volume} {86}},\ \bibinfo {pages} {1881} (\bibinfo {year}
  {2001})}\BibitemShut {NoStop}%
\bibitem [{\citenamefont {Chubukov}\ \emph {et~al.}(1994)\citenamefont
  {Chubukov}, \citenamefont {Senthil},\ and\ \citenamefont
  {Sachdev}}]{chubukov94}%
  \BibitemOpen
  \bibfield  {author} {\bibinfo {author} {\bibfnamefont {A.~V.}\ \bibnamefont
  {Chubukov}}, \bibinfo {author} {\bibfnamefont {T.}~\bibnamefont {Senthil}}, \
  and\ \bibinfo {author} {\bibfnamefont {S.}~\bibnamefont {Sachdev}},\ }\href
  {\doibase 10.1103/PhysRevLett.72.2089} {\bibfield  {journal} {\bibinfo
  {journal} {Phys. Rev. Lett.}\ }\textbf {\bibinfo {volume} {72}},\ \bibinfo
  {pages} {2089} (\bibinfo {year} {1994})}\BibitemShut {NoStop}%
\bibitem [{\citenamefont {Messio}\ \emph {et~al.}(2011)\citenamefont {Messio},
  \citenamefont {Lhuillier},\ and\ \citenamefont {Misguich}}]{Misguich11}%
  \BibitemOpen
  \bibfield  {author} {\bibinfo {author} {\bibfnamefont {L.}~\bibnamefont
  {Messio}}, \bibinfo {author} {\bibfnamefont {C.}~\bibnamefont {Lhuillier}}, \
  and\ \bibinfo {author} {\bibfnamefont {G.}~\bibnamefont {Misguich}},\ }\href
  {\doibase 10.1103/PhysRevB.83.184401} {\bibfield  {journal} {\bibinfo
  {journal} {Phys. Rev. B}\ }\textbf {\bibinfo {volume} {83}},\ \bibinfo
  {pages} {184401} (\bibinfo {year} {2011})}\BibitemShut {NoStop}%
\bibitem [{\citenamefont {Martin}\ and\ \citenamefont
  {Batista}(2008)}]{Martin2008}%
  \BibitemOpen
  \bibfield  {author} {\bibinfo {author} {\bibfnamefont {I.}~\bibnamefont
  {Martin}}\ and\ \bibinfo {author} {\bibfnamefont {C.~D.}\ \bibnamefont
  {Batista}},\ }\href {\doibase 10.1103/PhysRevLett.101.156402} {\bibfield
  {journal} {\bibinfo  {journal} {Phys. Rev. Lett.}\ }\textbf {\bibinfo
  {volume} {101}},\ \bibinfo {pages} {156402} (\bibinfo {year}
  {2008})}\BibitemShut {NoStop}%
\bibitem [{\citenamefont {Kato}\ \emph {et~al.}(2010)\citenamefont {Kato},
  \citenamefont {Martin},\ and\ \citenamefont {Batista}}]{Kato2010}%
  \BibitemOpen
  \bibfield  {author} {\bibinfo {author} {\bibfnamefont {Y.}~\bibnamefont
  {Kato}}, \bibinfo {author} {\bibfnamefont {I.}~\bibnamefont {Martin}}, \ and\
  \bibinfo {author} {\bibfnamefont {C.~D.}\ \bibnamefont {Batista}},\ }\href
  {\doibase 10.1103/PhysRevLett.105.266405} {\bibfield  {journal} {\bibinfo
  {journal} {Phys. Rev. Lett.}\ }\textbf {\bibinfo {volume} {105}},\ \bibinfo
  {pages} {266405} (\bibinfo {year} {2010})}\BibitemShut {NoStop}%
\bibitem [{\citenamefont {Rahmani}\ \emph {et~al.}(2013)\citenamefont
  {Rahmani}, \citenamefont {Muniz},\ and\ \citenamefont
  {Martin}}]{Rahmani2013}%
  \BibitemOpen
  \bibfield  {author} {\bibinfo {author} {\bibfnamefont {A.}~\bibnamefont
  {Rahmani}}, \bibinfo {author} {\bibfnamefont {R.~A.}\ \bibnamefont {Muniz}},
  \ and\ \bibinfo {author} {\bibfnamefont {I.}~\bibnamefont {Martin}},\ }\href
  {\doibase 10.1103/PhysRevX.3.031008} {\bibfield  {journal} {\bibinfo
  {journal} {Phys. Rev. X}\ }\textbf {\bibinfo {volume} {3}},\ \bibinfo {pages}
  {031008} (\bibinfo {year} {2013})}\BibitemShut {NoStop}%
\bibitem [{\citenamefont {Jiang}\ \emph {et~al.}(2014)\citenamefont {Jiang},
  \citenamefont {Mesaros},\ and\ \citenamefont {Ran}}]{YRan2014}%
  \BibitemOpen
  \bibfield  {author} {\bibinfo {author} {\bibfnamefont {S.}~\bibnamefont
  {Jiang}}, \bibinfo {author} {\bibfnamefont {A.}~\bibnamefont {Mesaros}}, \
  and\ \bibinfo {author} {\bibfnamefont {Y.}~\bibnamefont {Ran}},\ }\href
  {\doibase 10.1103/PhysRevX.4.031040} {\bibfield  {journal} {\bibinfo
  {journal} {Phys. Rev. X}\ }\textbf {\bibinfo {volume} {4}},\ \bibinfo {pages}
  {031040} (\bibinfo {year} {2014})}\BibitemShut {NoStop}%
\bibitem [{\citenamefont {Motrunich}(2006)}]{MotrunichOMF2006}%
  \BibitemOpen
  \bibfield  {author} {\bibinfo {author} {\bibfnamefont {O.~I.}\ \bibnamefont
  {Motrunich}},\ }\href {\doibase 10.1103/PhysRevB.73.155115} {\bibfield
  {journal} {\bibinfo  {journal} {Phys. Rev. B}\ }\textbf {\bibinfo {volume}
  {73}},\ \bibinfo {pages} {155115} (\bibinfo {year} {2006})}\BibitemShut
  {NoStop}%
\bibitem [{\citenamefont {Haldane}(1988)}]{Haldane1988}%
  \BibitemOpen
  \bibfield  {author} {\bibinfo {author} {\bibfnamefont {F.~D.~M.}\
  \bibnamefont {Haldane}},\ }\href {\doibase 10.1103/PhysRevLett.61.2015}
  {\bibfield  {journal} {\bibinfo  {journal} {Phys. Rev. Lett.}\ }\textbf
  {\bibinfo {volume} {61}},\ \bibinfo {pages} {2015} (\bibinfo {year}
  {1988})}\BibitemShut {NoStop}%
\bibitem [{\citenamefont {Jotzu}\ \emph {et~al.}(2014)\citenamefont {Jotzu},
  \citenamefont {Messer}, \citenamefont {Desbuquois}, \citenamefont {Lebrat},
  \citenamefont {Uehlinger}, \citenamefont {Greif},\ and\ \citenamefont
  {Esslinger}}]{Jotzu2014}%
  \BibitemOpen
  \bibfield  {author} {\bibinfo {author} {\bibfnamefont {G.}~\bibnamefont
  {Jotzu}}, \bibinfo {author} {\bibfnamefont {M.}~\bibnamefont {Messer}},
  \bibinfo {author} {\bibfnamefont {R.}~\bibnamefont {Desbuquois}}, \bibinfo
  {author} {\bibfnamefont {M.}~\bibnamefont {Lebrat}}, \bibinfo {author}
  {\bibfnamefont {T.}~\bibnamefont {Uehlinger}}, \bibinfo {author}
  {\bibfnamefont {D.}~\bibnamefont {Greif}}, \ and\ \bibinfo {author}
  {\bibfnamefont {T.}~\bibnamefont {Esslinger}},\ }\href
  {http://dx.doi.org/10.1038/nature13915} {\bibfield  {journal} {\bibinfo
  {journal} {Nature}\ }\textbf {\bibinfo {volume} {515}},\ \bibinfo {pages}
  {237} (\bibinfo {year} {2014})}\BibitemShut {NoStop}%
\bibitem [{\citenamefont {{Claassen}}\ \emph {et~al.}(2016)\citenamefont
  {{Claassen}}, \citenamefont {{Jiang}}, \citenamefont {{Moritz}},\ and\
  \citenamefont {{Devereaux}}}]{Devereaux2016}%
  \BibitemOpen
  \bibfield  {author} {\bibinfo {author} {\bibfnamefont {M.}~\bibnamefont
  {{Claassen}}}, \bibinfo {author} {\bibfnamefont {H.-C.}\ \bibnamefont
  {{Jiang}}}, \bibinfo {author} {\bibfnamefont {B.}~\bibnamefont {{Moritz}}}, \
  and\ \bibinfo {author} {\bibfnamefont {T.~P.}\ \bibnamefont {{Devereaux}}},\
  }\href@noop {} {\bibfield  {journal} {\bibinfo  {journal} {ArXiv e-prints}\ }
  (\bibinfo {year} {2016})},\ \Eprint {http://arxiv.org/abs/1611.07964}
  {arXiv:1611.07964 [cond-mat.str-el]} \BibitemShut {NoStop}%
\bibitem [{\citenamefont {{Kitamura}}\ \emph {et~al.}(2017)\citenamefont
  {{Kitamura}}, \citenamefont {{Oka}},\ and\ \citenamefont
  {{Aoki}}}]{Aoki2017}%
  \BibitemOpen
  \bibfield  {author} {\bibinfo {author} {\bibfnamefont {S.}~\bibnamefont
  {{Kitamura}}}, \bibinfo {author} {\bibfnamefont {T.}~\bibnamefont {{Oka}}}, \
  and\ \bibinfo {author} {\bibfnamefont {H.}~\bibnamefont {{Aoki}}},\
  }\href@noop {} {\bibfield  {journal} {\bibinfo  {journal} {ArXiv e-prints}\ }
  (\bibinfo {year} {2017})},\ \Eprint {http://arxiv.org/abs/1703.04315}
  {arXiv:1703.04315 [cond-mat.str-el]} \BibitemShut {NoStop}%
\bibitem [{\citenamefont {Bernu}\ \emph {et~al.}(1992)\citenamefont {Bernu},
  \citenamefont {Lhuillier},\ and\ \citenamefont {Pierre}}]{LhuillierQDJS1992}%
  \BibitemOpen
  \bibfield  {author} {\bibinfo {author} {\bibfnamefont {B.}~\bibnamefont
  {Bernu}}, \bibinfo {author} {\bibfnamefont {C.}~\bibnamefont {Lhuillier}}, \
  and\ \bibinfo {author} {\bibfnamefont {L.}~\bibnamefont {Pierre}},\ }\href
  {\doibase 10.1103/PhysRevLett.69.2590} {\bibfield  {journal} {\bibinfo
  {journal} {Phys. Rev. Lett.}\ }\textbf {\bibinfo {volume} {69}},\ \bibinfo
  {pages} {2590} (\bibinfo {year} {1992})}\BibitemShut {NoStop}%
\bibitem [{\citenamefont {Bernu}\ \emph {et~al.}(1994)\citenamefont {Bernu},
  \citenamefont {Lecheminant}, \citenamefont {Lhuillier},\ and\ \citenamefont
  {Pierre}}]{LhuillierSpectra1994}%
  \BibitemOpen
  \bibfield  {author} {\bibinfo {author} {\bibfnamefont {B.}~\bibnamefont
  {Bernu}}, \bibinfo {author} {\bibfnamefont {P.}~\bibnamefont {Lecheminant}},
  \bibinfo {author} {\bibfnamefont {C.}~\bibnamefont {Lhuillier}}, \ and\
  \bibinfo {author} {\bibfnamefont {L.}~\bibnamefont {Pierre}},\ }\href
  {\doibase 10.1103/PhysRevB.50.10048} {\bibfield  {journal} {\bibinfo
  {journal} {Phys. Rev. B}\ }\textbf {\bibinfo {volume} {50}},\ \bibinfo
  {pages} {10048} (\bibinfo {year} {1994})}\BibitemShut {NoStop}%
\bibitem [{\citenamefont {Cincio}\ and\ \citenamefont
  {Vidal}(2013)}]{Cincio2013}%
  \BibitemOpen
  \bibfield  {author} {\bibinfo {author} {\bibfnamefont {L.}~\bibnamefont
  {Cincio}}\ and\ \bibinfo {author} {\bibfnamefont {G.}~\bibnamefont {Vidal}},\
  }\href {\doibase 10.1103/PhysRevLett.110.067208} {\bibfield  {journal}
  {\bibinfo  {journal} {Phys. Rev. Lett.}\ }\textbf {\bibinfo {volume} {110}},\
  \bibinfo {pages} {067208} (\bibinfo {year} {2013})}\BibitemShut {NoStop}%
\bibitem [{\citenamefont {Zanardi}\ and\ \citenamefont
  {Paunkovi\ifmmode~\acute{c}\else \'{c}\fi{}}(2006)}]{Fidelity2006}%
  \BibitemOpen
  \bibfield  {author} {\bibinfo {author} {\bibfnamefont {P.}~\bibnamefont
  {Zanardi}}\ and\ \bibinfo {author} {\bibfnamefont {N.}~\bibnamefont
  {Paunkovi\ifmmode~\acute{c}\else \'{c}\fi{}}},\ }\href {\doibase
  10.1103/PhysRevE.74.031123} {\bibfield  {journal} {\bibinfo  {journal} {Phys.
  Rev. E}\ }\textbf {\bibinfo {volume} {74}},\ \bibinfo {pages} {031123}
  (\bibinfo {year} {2006})}\BibitemShut {NoStop}%
\bibitem [{\citenamefont {Gu}(2010)}]{FidelityReview2010}%
  \BibitemOpen
  \bibfield  {author} {\bibinfo {author} {\bibfnamefont {S.-J.}\ \bibnamefont
  {Gu}},\ }\href {\doibase 10.1142/S0217979210056335} {\bibfield  {journal}
  {\bibinfo  {journal} {International Journal of Modern Physics B}\ }\textbf
  {\bibinfo {volume} {24}},\ \bibinfo {pages} {4371} (\bibinfo {year}
  {2010})}\BibitemShut {NoStop}%
\bibitem [{\citenamefont {{Thomson}}\ and\ \citenamefont
  {{Sachdev}}(2016)}]{AlexTorus2016}%
  \BibitemOpen
  \bibfield  {author} {\bibinfo {author} {\bibfnamefont {A.}~\bibnamefont
  {{Thomson}}}\ and\ \bibinfo {author} {\bibfnamefont {S.}~\bibnamefont
  {{Sachdev}}},\ }\href@noop {} {\bibfield  {journal} {\bibinfo  {journal}
  {ArXiv e-prints}\ } (\bibinfo {year} {2016})},\ \Eprint
  {http://arxiv.org/abs/1607.05279} {arXiv:1607.05279 [cond-mat.str-el]}
  \BibitemShut {NoStop}%
\bibitem [{\citenamefont {Machida}\ \emph {et~al.}(2010)\citenamefont
  {Machida}, \citenamefont {Nakatsuji}, \citenamefont {Onoda}, \citenamefont
  {Tayama},\ and\ \citenamefont {Sakakibara}}]{Nakatsuji2010}%
  \BibitemOpen
  \bibfield  {author} {\bibinfo {author} {\bibfnamefont {Y.}~\bibnamefont
  {Machida}}, \bibinfo {author} {\bibfnamefont {S.}~\bibnamefont {Nakatsuji}},
  \bibinfo {author} {\bibfnamefont {S.}~\bibnamefont {Onoda}}, \bibinfo
  {author} {\bibfnamefont {T.}~\bibnamefont {Tayama}}, \ and\ \bibinfo {author}
  {\bibfnamefont {T.}~\bibnamefont {Sakakibara}},\ }\href
  {http://dx.doi.org/10.1038/nature08680} {\bibfield  {journal} {\bibinfo
  {journal} {Nature}\ }\textbf {\bibinfo {volume} {463}},\ \bibinfo {pages}
  {210} (\bibinfo {year} {2010})}\BibitemShut {NoStop}%
\bibitem [{\citenamefont {Greiter}\ and\ \citenamefont
  {Thomale}(2009)}]{GreiterNACSL2009}%
  \BibitemOpen
  \bibfield  {author} {\bibinfo {author} {\bibfnamefont {M.}~\bibnamefont
  {Greiter}}\ and\ \bibinfo {author} {\bibfnamefont {R.}~\bibnamefont
  {Thomale}},\ }\href {\doibase 10.1103/PhysRevLett.102.207203} {\bibfield
  {journal} {\bibinfo  {journal} {Phys. Rev. Lett.}\ }\textbf {\bibinfo
  {volume} {102}},\ \bibinfo {pages} {207203} (\bibinfo {year}
  {2009})}\BibitemShut {NoStop}%
\bibitem [{\citenamefont {Scharfenberger}\ \emph {et~al.}(2011)\citenamefont
  {Scharfenberger}, \citenamefont {Thomale},\ and\ \citenamefont
  {Greiter}}]{GreiterNACSL2011}%
  \BibitemOpen
  \bibfield  {author} {\bibinfo {author} {\bibfnamefont {B.}~\bibnamefont
  {Scharfenberger}}, \bibinfo {author} {\bibfnamefont {R.}~\bibnamefont
  {Thomale}}, \ and\ \bibinfo {author} {\bibfnamefont {M.}~\bibnamefont
  {Greiter}},\ }\href {\doibase 10.1103/PhysRevB.84.140404} {\bibfield
  {journal} {\bibinfo  {journal} {Phys. Rev. B}\ }\textbf {\bibinfo {volume}
  {84}},\ \bibinfo {pages} {140404} (\bibinfo {year} {2011})}\BibitemShut
  {NoStop}%
\bibitem [{\citenamefont {Greiter}\ \emph
  {et~al.}(2014{\natexlab{b}})\citenamefont {Greiter}, \citenamefont
  {Schroeter},\ and\ \citenamefont {Thomale}}]{GreiterNACSL2014}%
  \BibitemOpen
  \bibfield  {author} {\bibinfo {author} {\bibfnamefont {M.}~\bibnamefont
  {Greiter}}, \bibinfo {author} {\bibfnamefont {D.~F.}\ \bibnamefont
  {Schroeter}}, \ and\ \bibinfo {author} {\bibfnamefont {R.}~\bibnamefont
  {Thomale}},\ }\href {\doibase 10.1103/PhysRevB.89.165125} {\bibfield
  {journal} {\bibinfo  {journal} {Phys. Rev. B}\ }\textbf {\bibinfo {volume}
  {89}},\ \bibinfo {pages} {165125} (\bibinfo {year}
  {2014}{\natexlab{b}})}\BibitemShut {NoStop}%
\bibitem [{\citenamefont {{Liu}}\ \emph {et~al.}(2015)\citenamefont {{Liu}},
  \citenamefont {{Tu}}, \citenamefont {{Wu}}, \citenamefont {{He}},
  \citenamefont {{Liu}}, \citenamefont {{Zhou}},\ and\ \citenamefont
  {{Ng}}}]{TKNg2015}%
  \BibitemOpen
  \bibfield  {author} {\bibinfo {author} {\bibfnamefont {Z.-X.}\ \bibnamefont
  {{Liu}}}, \bibinfo {author} {\bibfnamefont {H.-H.}\ \bibnamefont {{Tu}}},
  \bibinfo {author} {\bibfnamefont {Y.-H.}\ \bibnamefont {{Wu}}}, \bibinfo
  {author} {\bibfnamefont {R.-Q.}\ \bibnamefont {{He}}}, \bibinfo {author}
  {\bibfnamefont {X.-J.}\ \bibnamefont {{Liu}}}, \bibinfo {author}
  {\bibfnamefont {Y.}~\bibnamefont {{Zhou}}}, \ and\ \bibinfo {author}
  {\bibfnamefont {T.-K.}\ \bibnamefont {{Ng}}},\ }\href@noop {} {\bibfield
  {journal} {\bibinfo  {journal} {ArXiv e-prints}\ } (\bibinfo {year}
  {2015})},\ \Eprint {http://arxiv.org/abs/1509.00391} {arXiv:1509.00391
  [cond-mat.str-el]} \BibitemShut {NoStop}%
\bibitem [{\citenamefont {Lecheminant}\ and\ \citenamefont
  {Tsvelik}(2017)}]{Tsvelik2017}%
  \BibitemOpen
  \bibfield  {author} {\bibinfo {author} {\bibfnamefont {P.}~\bibnamefont
  {Lecheminant}}\ and\ \bibinfo {author} {\bibfnamefont {A.~M.}\ \bibnamefont
  {Tsvelik}},\ }\href {\doibase 10.1103/PhysRevB.95.140406} {\bibfield
  {journal} {\bibinfo  {journal} {Phys. Rev. B}\ }\textbf {\bibinfo {volume}
  {95}},\ \bibinfo {pages} {140406} (\bibinfo {year} {2017})}\BibitemShut
  {NoStop}%
\bibitem [{\citenamefont {{Saadatmand}}\ and\ \citenamefont
  {{McCulloch}}(2017)}]{McCulloch2017}%
  \BibitemOpen
  \bibfield  {author} {\bibinfo {author} {\bibfnamefont {S.~N.}\ \bibnamefont
  {{Saadatmand}}}\ and\ \bibinfo {author} {\bibfnamefont {I.~P.}\ \bibnamefont
  {{McCulloch}}},\ }\href@noop {} {\bibfield  {journal} {\bibinfo  {journal}
  {ArXiv e-prints}\ } (\bibinfo {year} {2017})},\ \Eprint
  {http://arxiv.org/abs/1704.03418} {arXiv:1704.03418 [cond-mat.str-el]}
  \BibitemShut {NoStop}%
\bibitem [{\citenamefont {{Gong}}\ \emph {et~al.}(2017)\citenamefont {{Gong}},
  \citenamefont {{Zhu}}, \citenamefont {{Zhu}}, \citenamefont {{Sheng}},\ and\
  \citenamefont {{Yang}}}]{ShengDMRG2017}%
  \BibitemOpen
  \bibfield  {author} {\bibinfo {author} {\bibfnamefont {S.-S.}\ \bibnamefont
  {{Gong}}}, \bibinfo {author} {\bibfnamefont {W.}~\bibnamefont {{Zhu}}},
  \bibinfo {author} {\bibfnamefont {J.-X.}\ \bibnamefont {{Zhu}}}, \bibinfo
  {author} {\bibfnamefont {D.~N.}\ \bibnamefont {{Sheng}}}, \ and\ \bibinfo
  {author} {\bibfnamefont {K.}~\bibnamefont {{Yang}}},\ }\href@noop {}
  {\bibfield  {journal} {\bibinfo  {journal} {ArXiv e-prints}\ } (\bibinfo
  {year} {2017})},\ \Eprint {http://arxiv.org/abs/1705.00510} {arXiv:1705.00510
  [cond-mat.str-el]} \BibitemShut {NoStop}%
\end{thebibliography}

%

\end{document}